\pdfoutput=1
\documentclass[11pt,twoside,a4paper,cmspaper,final,collab]{cms-tdr}
\def\svnVersion{bc92dc8-D}\def\svnDate{2021/10/11}\def\cmsCernNoTag{CERN-EP-2021-095}\def\cmsCernDate{\today}\def\cmsMessage{"Published in Physical Review D as \href{http://dx.doi.org/10.1103/PhysRevD.104.072001}{\doi{10.1103/PhysRevD.104.072001}.}"}
\begin{document}\cmsNoteHeader{SMP-20-016}

\newlength\cmsFigWidth
\newlength\cmsTabSkip\setlength{\cmsTabSkip}{1.7ex}
\ifthenelse{\boolean{cms@external}}{\setlength\cmsFigWidth{\columnwidth}}{\setlength\cmsFigWidth{0.6\textwidth}}
\ifthenelse{\boolean{cms@external}}{\providecommand{\cmsTable}[1]{#1}}{\providecommand{\cmsTable}[1]{\resizebox{\textwidth}{!}{#1}}}
\newcommand{\mll}{m_{\ell\ell}}
\newcommand{\mjj}{m_\mathrm{jj}}
\newcommand{\mzg}{m_{\PZ\PGg}}
\newcommand{\etajj}{\abs{\Delta \eta_{\mathrm{jj}}}}
\newcommand{\gbarrel}{\gamma_{\text{barrel}}}
\newcommand{\gendcap}{\gamma_{\text{endcap}}}
\newcommand{\dphizgjj}{\Delta \phi_{\PZ\PGg, \mathrm{jj}}}

\cmsNoteHeader{SMP-20-016}

\title{Measurement of the electroweak production of \texorpdfstring{$\PZ\PGg$}{Zgamma} and two jets in proton-proton collisions at \texorpdfstring{$\sqrt{s}=13\TeV$}{sqrt(s) = 13 TeV} and constraints on anomalous quartic gauge couplings}

\date{\today}

\abstract{The first observation of the electroweak (EW) production of a {\PZ} boson, a photon, and two forward jets ($\PZ\PGg$jj) in proton-proton collisions at a center-of-mass energy of 13\TeV is presented. A data set corresponding to an integrated luminosity of 137\fbinv, collected by the CMS experiment at the LHC in 2016--2018 is used. The measured fiducial cross section for EW $\PZ\PGg$jj is $\sigma_{\mathrm{EW}}=5.21\pm0.52\stat\pm0.56\syst\unit{fb}=5.21\pm0.76\unit{fb}$. Single-differential cross sections in photon, leading lepton, and leading jet transverse momenta, and double-differential cross sections in $m_{\mathrm{jj}}$ and $\abs{\Delta\eta_{\mathrm{jj}}}$ are also measured. Exclusion limits on anomalous quartic gauge couplings are derived at 95\% confidence level in terms of the effective field theory operators $\mathrm{M}_{0}$ to $\mathrm{M}_{5}$, $\mathrm{M}_{7}$, $\mathrm{T}_{0}$ to $\mathrm{T}_{2}$, and $\mathrm{T}_{5}$ to $\mathrm{T}_{9}$.}

\hypersetup{
pdfauthor={CMS Collaboration},
pdftitle={Measurement of the electroweak production of Zgamma and two jets in proton-proton collisions at sqrt(s) = 13 TeV and constraints on anomalous quartic gauge couplings},
pdfsubject={CMS},
pdfkeywords={CMS, vector boson scattering}}

\maketitle 
\section{Introduction}
\label{intro}
Vector boson scattering (VBS) processes are purely electroweak (EW) interactions at leading order (LO). In a proton-proton ($\Pp\Pp$) collision where two vector bosons radiated from the incoming quarks scatter, the two outgoing quarks appear as jets widely separated in pseudorapidity ($\eta$) and with a large dijet mass ($\mjj$), providing a unique signature. The VBS is of great interest because of the role of the Higgs boson in restoring unitarity to the VBS cross section. Studies of VBS complement direct Higgs boson measurements~\cite{Aad:2012tfa,Chatrchyan:2012ufa,Chatrchyan:2012ufa_long,higgs_measure_2016,higgs_measure_2019} and open a window to beyond the standard model (BSM) scenarios at energy scales outside the reach of direct searches. Standard model (SM) EW production of $\PZ\PGg$jj, which can proceed through VBS, can be extracted by exploiting the unique features of the VBS signature. A precise measurement of EW $\PZ\PGg$jj production is sensitive not only to quartic gauge couplings (QGCs) in the SM as well as possible anomalous QGCs (aQGCs)~\cite{Eboli:2006wa}, but also to triple gauge couplings (TGCs) and anomalous TGCs. Since the latter is well constrained in diboson production~\cite{aTGC_WV}, they are not explored in this paper.

\begin{figure*}[!ht]
\centering
      \includegraphics[width=0.32\textwidth]{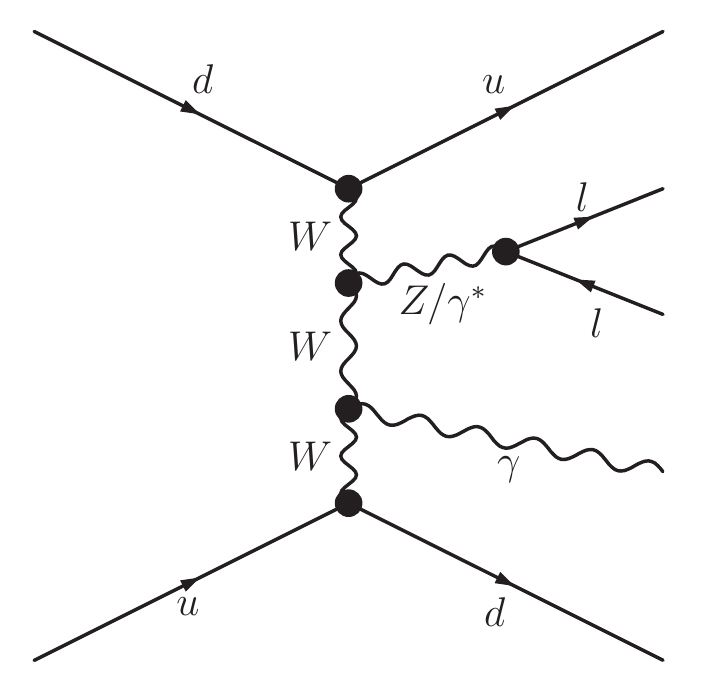}
      \includegraphics[width=0.32\textwidth]{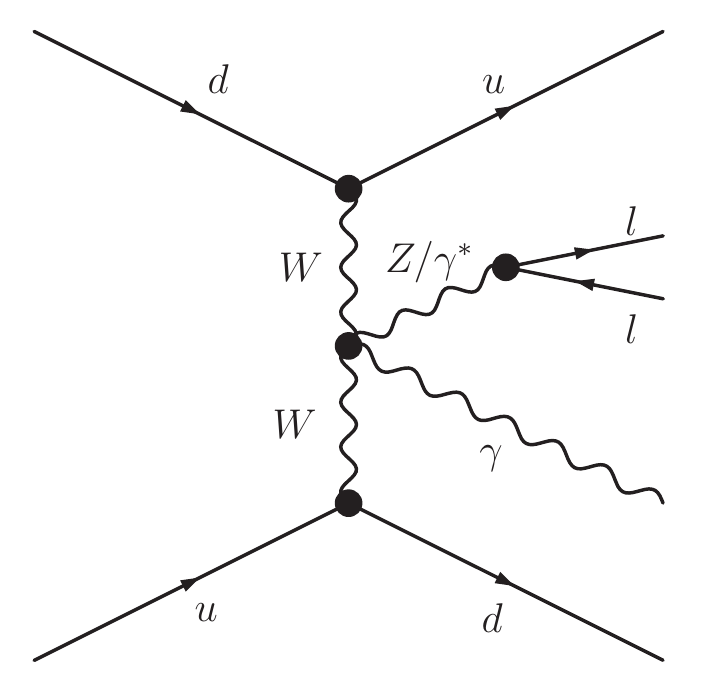}
      \includegraphics[width=0.32\textwidth]{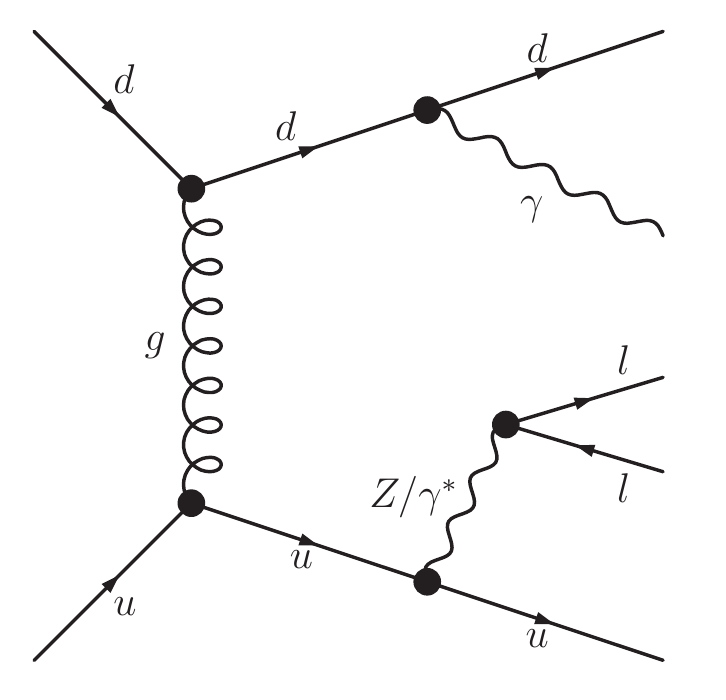}\\     
      \includegraphics[width=0.32\textwidth]{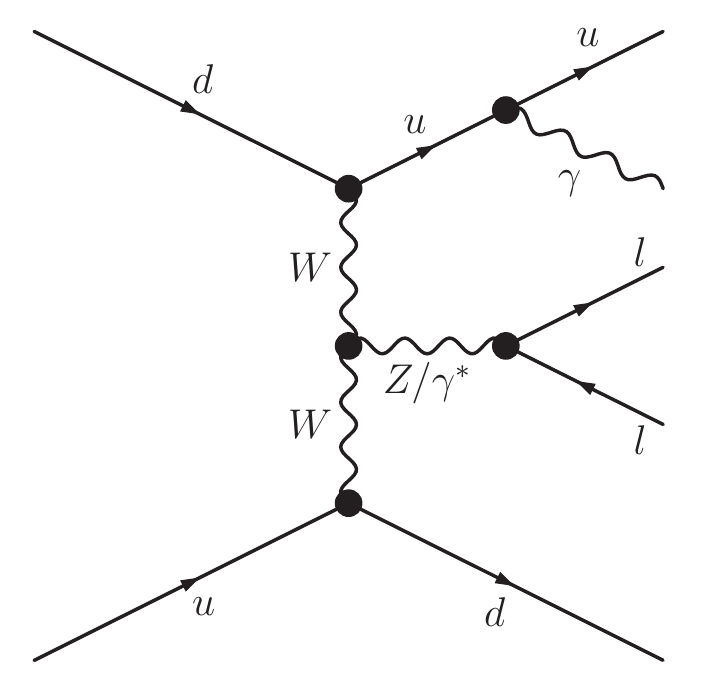}
      \includegraphics[width=0.32\textwidth]{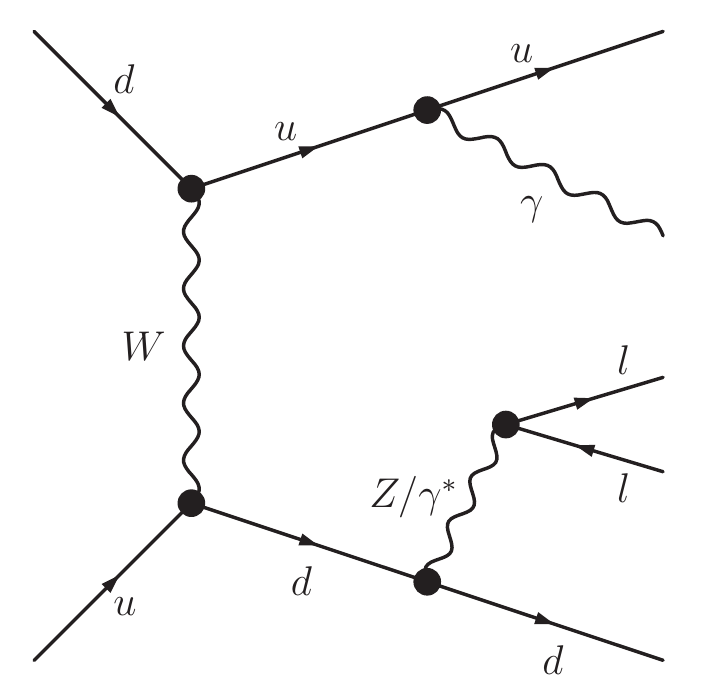}
      \includegraphics[width=0.32\textwidth]{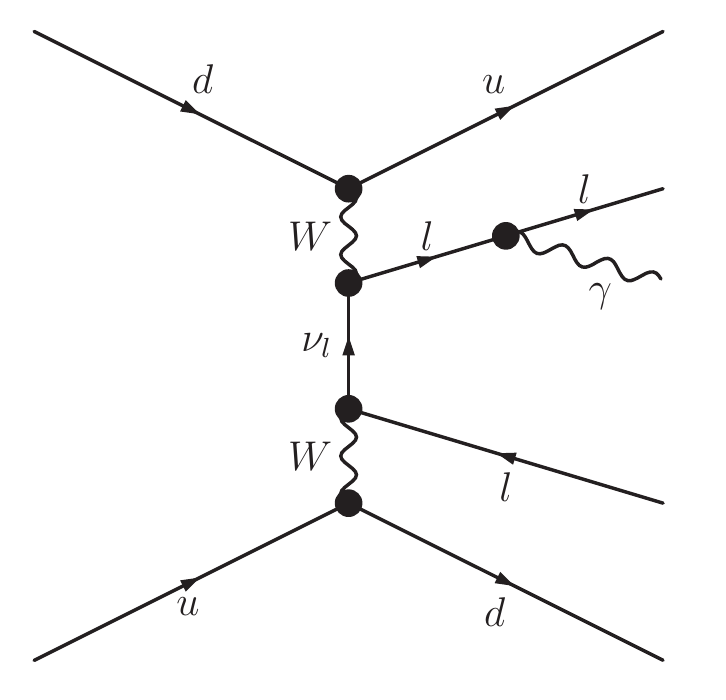}\\
\caption{Representative Feynman diagrams for $\PZ\PGg$jj production. With the exception of the upper right one, the diagrams involve only EW vertices: VBS via {\PW} boson~(upper left), VBS with QGC~(upper center), vector boson fusion with TGCs~(lower left), bremsstrahlung~(lower center), multiperipheral~(lower right), whereas the upper right diagram represents a QCD-induced contribution.}
\label{fig:za_feynman}
\end{figure*}

In this paper, we present the first observation of EW $\PZ\PGg$jj production. Events corresponding to the $\ell^{+}\ell^{-}\PGg$jj final states where $\ell=\Pe$ or $\mu$, are selected, and the dijet system is required to satisfy the typical VBS signature. Figure~\ref{fig:za_feynman} shows the representative Feynman diagrams (upper left and center), in which a {\PZ} boson and a photon are produced in a scattering interaction between two $\PW$ bosons, an example of the production of $\PZ\PGg$jj induced by quantum chromodynamics (upper right), one of the important backgrounds in this study, and examples of non-VBS EW production of $\PZ\PGg$jj (lower).

Previous experimental results for EW $\PZ\PGg$jj production have been reported by the ATLAS~\cite{Aad:2019wpb} and CMS~\cite{Sirunyan:2020tlu} Collaborations based on data collected in 2016 at $\sqrt{s}=13\TeV$, corresponding to integrated luminosities of 35.9\fbinv  and 36\fbinv , respectively. The observed (expected) significance reported by ATLAS was 4.1 (4.1) standard deviations (SD), and by CMS was 4.7 (5.5) SD. 

The measurements reported here are based on $\Pp\Pp$ collision data at $\sqrt{s}=13\TeV$, collected from 2016 to 2018 by the CMS experiment at the CERN LHC, corresponding to an integrated luminosity of 137\fbinv . A simultaneous maximum likelihood (ML) fit is used to extract the signal significance, the signal strength, fiducial cross sections, and unfolded differential cross sections for both EW and EW+QCD production of $\PZ\PGg$jj. Unfolded differential cross sections are measured as functions of three 1-dimensional observables (the transverse momenta (\pt) of the leading lepton, photon, and jet) and one 2-dimensional observable ($m_{\mathrm{jj}}$ and $\abs{\Delta\eta_{\mathrm{jj}}}$). Using the selected $\ell^{+}\ell^{-}\PGg$jj events with high photon \pt, constraints on BSM contributions to the $\PV\PV\PZ\PGg$ vertices, where $\PV=\PW$, {\PZ}, or {\PGg}, are determined in an effective field theory (EFT) framework using dimension-8 operators~\cite{Eboli:2006wa}. 

This paper is arranged as follows. The CMS detector and the event simulation are summarized in Secs~\ref{detector} and \ref{sec:samples}. The object reconstruction and the event selection are described in Sec~\ref{reco_and_selection}. The estimation of the main backgrounds is given in Sec~\ref{sec:bkg_estimation}. The systematic uncertainties are discussed in Sec~\ref{sec:uncertainties} and the results are presented in Sec~\ref{results}. The paper is summarized in Sec~\ref{sec:summary}.

\section{The CMS detector}
\label{detector}
The central feature of the CMS~\cite{Chatrchyan:2008zzk} apparatus is a superconducting solenoid of 6\unit{m} internal diameter, providing a magnetic field of 3.8\unit{T}. Within the solenoid volume are a silicon pixel and strip tracker, a lead tungstate crystal electromagnetic calorimeter (ECAL), and a brass and scintillator hadron calorimeter (HCAL), each composed of a barrel and two endcap sections. Forward calorimeters extend the $\eta$ coverage provided by the barrel and endcap detectors. Muons are detected in gas-ionization chambers embedded in the steel flux-return yoke outside the solenoid.

Events of interest are selected using a two-level trigger system~\cite{trigger_system}. The first level (L1), composed of custom hardware processors, uses information from the calorimeters and muon detectors to select events of interest with a maximum rate of 100\unit{kHz} within a latency of less than 4\mus. The second level is a high-level trigger (HLT) processor, made up of a farm of processors running a version of the full event reconstruction software optimized for fast processing, and decreases the rate to about 1\unit{kHz} before storage.

A more detailed description of the CMS detector, together with a definition of the coordinate system and kinematic variables, is reported in Ref.~\cite{Chatrchyan:2008zzk}.

\section{Signal and background simulation}
\label{sec:samples}
Samples of simulated events are used to model the EW $\PZ\PGg$jj signal and a variety of background processes. Since the data analyzed for this paper were collected by the CMS experiment from 2016 through 2018, the Monte Carlo (MC) samples are also simulated separately for each year with different versions of the MC generators. The \MGvATNLO~\cite{MGatNLO} v2.6.0, which is abbreviated as MG5 in the following text, \POWHEG~\cite{POWHEG,POWHEG1,POWHEG2,POWHEG3} v2.0, and \PYTHIA~\cite{Sjostrand:2014zea} v8.226 are used for 2016. MG5 v2.6.1, \POWHEG v2.0, and \PYTHIA v8.230 are used for 2017 and 2018. The simulated samples are used to establish the event selection, optimize the signal extraction procedure, and estimate the total signal efficiency in the inclusive fiducial volume and single signal efficiency in every bin of final experimental distributions. 

The signal EW $\PZ\PGg$jj production with the {\PZ} boson decaying to a pair of leptons is simulated at LO using the MG5 generator. The main backgrounds, including QCD-induced $\PZ\PGg$ (QCD $\PZ\PGg$) and {\ttbar}$\gamma$ (TT$\gamma$) are generated with MG5 with up to one jet at next-to-leading order (NLO) in QCD using the FxFx jet merging scheme~\cite{Frederix:2012ps}. The remaining backgrounds, including diboson ({\PW}{\PW}/{\PW}{\PZ}/{\PZ}{\PZ}) production ($\PV\PV$) and single top quark production (ST), are generated with either \PYTHIA at LO or \POWHEG at NLO. The interference between EW $\PZ\PGg$jj production and QCD-induced $\PZ\PGg$jj production is also simulated with MG5 at $\mathcal{O}(\alpS\alpha_{\mathrm{EW}}^{3})$ and treated as a part of QCD-induced $\PZ\PGg$.

The simulation of aQGCs is performed using MG5 at LO. The matrix element reweighting functionality~\cite{mg5:reweight} is employed to produce additional weights that correspond to an appropriately spaced grid for each of the anomalous couplings probed in this paper.

The protons are described in the simulation by the NNPDF~3.0~\cite{nnpdf} (NNPDF~3.1~\cite{nnpdf_new}) parton distribution functions (PDFs) for 2016 (2017 and 2018). The \PYTHIA generator package is used for the proton showering, hadronization, and underlying event simulation. The samples are interfaced to \PYTHIA with the tune CUETP8M1 (CP5)~\cite{Skands:2014pea,Khachatryan:2015pea} for the year 2016 (2017 and 2018).

The detector response is modeled via a detailed description of the CMS detector implemented in the \GEANTfour~package~\cite{geant4, geant4_2}. The simulated events are reconstructed in the same way as the CMS data and include additional interactions in the same and neighboring bunch crossings, referred to as pileup (PU). Simulated PU events are weighted so that the number of reconstructed primary vertices reproduces that observed in the data.

\section{Object reconstruction and event selection}
\label{reco_and_selection}
\subsection{Object reconstruction}
\label{sec:event rec}
The final state of interest consists of a pair of oppositely charged isolated leptons, a photon, and two jets. We employ both a global object reconstruction algorithm and dedicated particle-type (\eg, muon) reconstruction algorithms, for different purposes, as described in the following paragraphs. The global object reconstruction algorithm, called the particle-flow (PF) algorithm~\cite{Sirunyan:2017ulk}, reconstructs and identifies all particle candidates (photons, electrons, muons, charged and neutral hadrons, and missing transverse momentum) in an event, based on a combination of information from the various elements of the CMS detector; the result is a set of physics objects called PF candidates.

The candidate vertex with the largest value of summed physics-object $\pt^2$ is taken to be the primary $\Pp\Pp$ interaction vertex. The physics objects are the jets, clustered using the jet finding algorithm~\cite{antikt,Cacciari:fastjet1} with the tracks assigned to candidate vertices as inputs, and the associated missing transverse momentum, taken as the negative vector sum of the \pt of those jets, which includes the leptons.

Electron candidates are reconstructed by combining information from the ECAL and the tracker within $\abs{\eta}<2.5$ and $\pt>25\GeV$. The energy of electrons is determined from a combination of the electron momentum at the primary interaction vertex determined by the tracker, the energy of the corresponding ECAL cluster, and the energy sum of all bremsstrahlung photons spatially compatible with originating from the electron track. Reconstructed electrons are required to satisfy additional identification requirements~\cite{cmscollaboration2020electron} as follows: the relative amount of energy deposited in the HCAL; a matching procedure between the trajectory in the inner tracker and that in the supercluster~\cite{Khachatryan:2015hwa} of the ECAL; the number of missing hits in the inner tracker; the consistency between the track and primary vertex; a shower shape variable $\sigma_{\mathrm{i}\eta\mathrm{i}\eta}$, which quantifies the transverse spread in $\eta$ of the electromagnetic shower in the ECAL (discussed in Sec~\ref{sec:bkg_estimation}); and a photon conversion rejection algorithm. An appropriate working point, referred to as the stringent electron selection, which has an average per-electron efficiency of 80\%, is used to identify the electron candidates from the \PZ boson decays in the signal process. A far less restrictive working point, referred to as the minimal electron selection, which has an average per-electron efficiency of 95\%, is used to remove events that contain additional electrons.

Muon candidates are reconstructed by combining information from the silicon tracker and the muon system within the region $\abs{\eta}<2.4$ and $\pt>20\GeV$~\cite{muon_13tev}. The combined information is used to produce a global track fit, and the muon momenta are obtained from the track curvatures. Muons are selected from the reconstructed muon track candidates by applying additional identification requirements as follows: the number of hits in the muon system and the inner tracker; the quality of the combined fit to a track; the number of matched muon-detector planes; and the consistency between the track and primary vertex. Different muon identification working points are defined according to their efficiency. An appropriate working point, referred to as the stringent muon selection, is used to identify the muon candidates from the \PZ boson decays in the signal process. The efficiency to reconstruct and identify muons is greater than 96\%. A far less restrictive working point, referred to as the minimal muon selection, is used to remove events with additional muons.

Leptons are required to be isolated from other particles in the event. The relative isolation is used and defined as:
\begin{linenomath}
   \ifthenelse{\boolean{cms@external}}{
\begin{multline}\label{iso}
R_\text{Iso} = \Bigl(\sum \pt^{\text{charged}} \\
+ \text{max}\left[0, \sum \pt^{\text{neutral}} + \sum \pt^{\PGg} - \pt^{\mathrm{PU}}\right]\Bigr)\Big/\pt^{\ell},
\end{multline}
   }{
      \begin{equation}\label{iso}
         R_\text{Iso} = \Bigl(\sum \pt^{\text{charged}} 
         + \text{max}\left[0, \sum \pt^{\text{neutral}} + \sum \pt^{\PGg} - \pt^{\mathrm{PU}}\right]\Bigr)\Big/\pt^{\ell},
         \end{equation}
   }
\end{linenomath}
where the sums run over the charged and neutral hadrons, as well as the photons, in a cone of size $\Delta R = \sqrt{\smash[b]{(\Delta\eta)^2+(\Delta\phi)^2}} = 0.3$ (0.4) for the electron (muon) trajectory, where $\eta$ and $\phi$ denote the pseudorapidity and azimuthal angle. The quantity $\sum \pt^{\text{charged}}$ is the scalar \pt sum of charged hadrons originating from the primary vertex; $\sum \pt^{\text{neutral}}$ and $\sum \pt^{\PGg}$ are the respective scalar \pt sums of neutral hadrons and photons. The contribution from PU in the isolation cone, $\pt^{\mathrm{PU}}$, is subtracted using the \FASTJET v3.0.2 technique~\cite{Cacciari:fastjet1}. For electrons, $\pt^{\mathrm{PU}}$ is evaluated using the ``jet area'' method described in Ref.~\cite{jetarea_method}. For muons, $\pt^{\mathrm{PU}}$ is assumed to be half of the scalar \pt sum deposited in the isolation cone by charged particles not associated with the primary vertex. The factor of one half corresponds approximately to the ratio of neutral to charged hadrons produced in the hadronization of PU interactions. Electrons are considered isolated for the stringent (minimal) working points if $R_{\text{Iso}}<0.0695$ ($0.1750$) in the barrel and $R_{\text{Iso}}<0.0821$ (0.1590) in the endcap detector regions. Muons are considered isolated for the stringent (minimal) working points if $R_{\text{Iso}}<0.15$ (0.25).

The efficiencies of lepton reconstruction and selection are measured as a function of $\pt^{\ell}$ and $\eta_{\ell}$ for both data events and MC events. The ``tag-and-probe'' technique~\cite{tagandprobe} is used on events containing a single \PZ boson. The ratio of efficiencies from events in data and MC are used to correct the simulation. The momentum scales of both muons and electrons are calibrated in bins of $\pt^{\ell}$ and $\eta_{\ell}$~\cite{cmscollaboration2020electron,electron_7tev}.

Photon reconstruction and selection are similar to electron reconstruction and selection. Photons located in the barrel region, $\abs{\eta}<1.442$, and the ECAL endcap region, $1.566<\abs{\eta}<2.500$, with $\pt>20\GeV$, are referred to as $\gbarrel$ and $\gendcap$, respectively. The region $1.442<\abs{\eta}<1.566$ is a transition region between the barrel and endcaps and is not used for photon reconstruction, because the reconstruction of a photon object in this region is less precise. Reconstructed photons are required to satisfy further quality criteria~\cite{cmscollaboration2020electron} based on the following quantities: the relative amount of deposited energy in the ECAL and HCAL; isolation variables constructed separately for the charged and neutral hadrons, photons other than the signal photon; and a procedure that quantifies the likelihood for a photon to originate from electron bremsstrahlung. An appropriate working point, referred to as the stringent photon selection, with an average per-photon efficiency of 80\%, is used to reconstruct prompt photons (not from hadron decays) in the final state. A second working point, which will be referred to as the nonprompt-enriched photon selection, is used to reconstruct nonprompt photons that are mainly products of neutral pion and $\eta$ meson decays~\cite{photon_8tev,cmscollaboration2020electron}, which constitute an important background in this study. 

Jets are reconstructed from particle-flow candidates using the anti-\kt jet clustering algorithm~\cite{antikt}, as implemented in the \FASTJET package v3.0.2, with a distance parameter of 0.4. The energies of charged hadrons are determined from a combination of their momenta measured in the tracker and the matching of ECAL and HCAL energy deposits, corrected for the response of the calorimeters to hadronic showers. The energy of neutral hadrons is obtained from the corresponding corrected ECAL and HCAL energies. To reduce the instrumental background, as well as the contamination from PU, jets are selected by a stringent jet selection based on the multiplicities and energy fractions carried by charged and neutral hadrons.

Jet energy corrections are extracted from data and simulated events to account for the effects of PU, non-uniformity of the detector response, and residual differences between the jet energy scale (JES) in data and simulation. The JES calibration~\cite{jer} relies on corrections parametrized in terms of the uncorrected \pt and $\eta$ of the jet, and is applied as a multiplicative factor, scaling the four-momentum vector of each jet. To ensure that jets are well measured and to reduce the PU contamination, all jets must satisfy $\abs{\eta}<4.7$ and have a corrected $\pt>30\GeV$. Jets from PU are further rejected using PU jet identification criteria based on a multivariate technique~\cite{puId}.

\subsection{Event selection}
\label{sec:event_selection}

The events of interest are selected by dilepton triggers. We denote the lepton having the larger \pt as $\ell1$ and the other one as $\ell2$. The \pt thresholds in the HLT are $\pt^{\ell1}>23\GeV$, $\pt^{\ell2}>12\GeV$ for electrons, and $\pt^{\ell1}>17\GeV$, and $\pt^{\ell2}>8\GeV$ for muons. In 2016 and 2017, a timing shift in the ECAL endcap was not properly accounted for in the trigger logic, resulting in the trigger decision sometimes mistakenly being assigned to the previous bunch crossing. This led to a sizable decrease in the L1 trigger efficiency for events with high energy deposits in the ECAL endcaps. The loss of efficiency for EW $\PZ\PGg$jj events associated with this effect is $\approx$8\% for the invariant mass of two jets $\mjj>500\GeV$, and increases to $\approx$15\% for $\mjj>2\TeV$. A correction is therefore applied as a function of jet \pt and $\eta$ using an unbiased data sample with correct timing and is implemented through a factor that represents the probability of the event to avoid having mistimed signals.

Selected events are required to contain two oppositely charged same-flavor leptons, either a pair of electrons or a pair of muons. Both leptons must pass the stringent working points and the corresponding isolation requirements described in Sec~\ref{sec:event rec} and must satisfy $\pt>25$ $(20)\GeV$, and $\abs{\eta}<2.5$ (2.4) in the electron (muon) case. The invariant mass of the dilepton system ($\mll$) is required to be within the window $70<\mll<110\GeV$. To reduce the {\PW}{\PZ} and {\PZ}{\PZ} backgrounds, events are rejected if there is any additional lepton passing the less restrictive identification criteria mentioned in Sec~\ref{sec:event rec}.  
At least one photon satisfying the stringent identification criteria is required. The photon with the largest \pt in the event is used if there is more than one photon passing the stringent identification criteria. The photon is required to have $\pt>20\GeV$. The $\Delta R$ between the selected photon and each of the selected leptons is required to be larger than $0.7$. The invariant mass of the dilepton-photon system must satisfy $m_{Z\gamma}>100\GeV$ to reduce the contribution from final-state radiation in \PZ boson decays. The event must also contain at least two jets that satisfy the jet identification criteria described in Sec~\ref{sec:event rec} and that are separated from selected leptons and photons by $\Delta R>0.5$. The jet with the largest \pt is referred to as the leading jet and is denoted as j1, and the jet with the lower \pt is denoted as j2. The jets are required to satisfy $\pt>30\GeV$, $\abs{\eta}<4.7$, and $\Delta R(\mathrm{j1,j2)}>0.5$. The dilepton selection with a photon and two jets is henceforth referred to as the ``common'' selection.

The signal region is defined by the common selection, and by requiring $m_{\mathrm{jj}}>500\GeV$ and $\etajj = \abs{\eta_{\mathrm{j1}} - \eta_{\mathrm{j2}}}>2.5$. Two additional criteria are used for the signal significance measurement. First, the Zeppenfeld variable~\cite{zeppenfeld} $\eta^* = \abs{\eta_{\PZ\PGg} - (\eta_{\mathrm{j1}} + \eta_{\mathrm{j2}})/2}$, where $\eta_{\PZ\PGg}$ represents the $\eta$ of the $\PZ\PGg$ system, needs to satisfy $\eta^*<2.4$. Second, the magnitude of the difference between the azimuthal angle of the $\PZ\PGg$ system ($\phi_{\PZ\gamma}$) and the azimuthal angle of the dijet system ($\phi_{\mathrm{j1j2}}$), $\dphizgjj=\abs{\phi_{\PZ\gamma}-\phi_{\mathrm{j1j2}}}$, which should be large in signal events because the two systems are recoiling against each other, must satisfy $\dphizgjj>1.9$. A low-$m_{\mathrm{jj}}$ control region, in which the EW signal is negligible compared with the contribution from QCD-induced $\PZ\PGg$jj production, is defined by the common selection and the requirement $150\GeV<m_{\mathrm{jj}}<500\GeV$.
 
The total and differential cross sections for EW $\PZ\PGg$jj and EW+QCD $\PZ\PGg$jj production are measured in a fiducial region (see ``Fiducial volume'' in Table~\ref{tab:selections}) that closely mirrors the EW signal in the VBS at the particle level, in which the requirements of $\eta^*$ and $\dphizgjj$ for the optimization of signal significance used in the EW signal region (see ``EW signal region'' in Table~\ref{tab:selections}) are not applied. The particle-level leptons and photons are required to be prompt, which means that the photon should be from the VBS process and lepton should be from the \PZ decay, and the momenta of prompt photons with $\Delta R_{\ell\gamma}<0.1$ are added to the lepton momenta to correct for final-state photon radiation.

The selection for the aQGC search is similar to the EW signal selection, but targets the characteristic high-energy behavior of aQGC processes by requiring $\pt^{\PGg}>120\GeV$. In this high photon $\pt$ region, the background has been suppressed as much as possible, so the requirements of $\eta^*$ and $\dphizgjj$ are not applied. 

A summary of all the selection criteria is displayed in Table~\ref{tab:selections}.

\begin{table*}[htbp]
\centering
\topcaption{Summary of the five sets of event selection criteria used to define events in the fiducial cross section measurement region, control region, EW signal extraction region, and the region used to search for aQGC contributions.}
\begin{scotch}{lc}
 Common selection & $\pt^{\ell1,\ell2}>25\GeV$, $\abs{\eta^{\ell1,\ell2}}<2.5$ for electron channel\\
              & $\pt^{\ell1,\ell2}>20\GeV$, $\abs{\eta^{\ell1,\ell2}}<2.4$ for muon channel\\
                  & $\pt^{\PGg}>20\GeV$, $\abs{\eta^{\PGg}}<1.442$ or $1.566<\abs{\eta^{\PGg}}<2.500$ \\
                  & $\pt^{\mathrm{j1,j2}}>30\GeV$, $\abs{\eta^{\mathrm{j1,j2}}}<4.7$\\
                  & $70<\mll<110\GeV$, $\mzg>100\GeV$\\
                  & $\Delta R_{\mathrm{jj}}$, $\Delta R_{\mathrm{j}\gamma}$, $\Delta R_{\mathrm{j}\ell}>0.5$, $\Delta R_{\ell\gamma}>0.7$\\[\cmsTabSkip]
Fiducial volume   & Common selection,\\
                  & $\mjj>500\GeV$, $\etajj>2.5$\\[\cmsTabSkip]  
Control region & Common selection,\\
               & $150<\mjj<500\GeV$ \\[\cmsTabSkip]
EW signal region  & Common selection,\\
                  &$\mjj>500\GeV$, $\etajj>2.5$,\\
                  & $\eta^*<2.4$, $\dphizgjj>1.9$\\[\cmsTabSkip]
aQGC search region      & Common selection,\\
                    & $\mjj>500\GeV$, $\etajj>2.5$,\\
                          & $\pt^{\PGg}>120\GeV$ \\
\end{scotch}
\label{tab:selections}
\end{table*}

\section{Background estimation}
\label{sec:bkg_estimation}
The dominant background arises from the QCD-induced production of $\PZ\PGg$jj. The yield and shape of this irreducible background are taken from simulation, but are ultimately constrained by the data in the ML fit mentioned in Sec~\ref{sec7} that extracts the EW signal. The second most important background arises from events in which the selected photon is not prompt and is mainly from {\PZ}+jets events. This background cannot be simulated accurately and is estimated from data, as described in the following paragraph. Other small contributions feature kinematic distributions similar to that of the dominant background and are estimated from simulation including single top quark events in the $s$- and $t$-channels that are normalized to their respective NLO cross sections; associated single top quark and {\PW} boson production normalized to its next-to-next-to-leading order (NNLO) cross section~\cite{STW_1}; {\PW}{\PW} production normalized to its NNLO cross section; {\PW}{\PZ}, {\PZ}{\PZ}, QCD-induced {\PW}$\PGg$jj, and {\ttbar}$\gamma$ production normalized to their NLO cross sections.

The background from events containing a nonprompt photon is estimated using data to calculate the event weight of the corresponding nonprompt photon event shown in Eq.~(\ref{fakerate}),
\begin{linenomath}
\begin{equation}\label{fakerate}
w(\pt^{\PGg})=\frac{n_{\text{data}}(\pt^{\PGg})}{N^{\text{unweighted}}_{\text{fake}}(\pt^{\PGg})}\epsilon_{\text{fake-fraction}}(\pt^{\PGg}),
\end{equation}
\end{linenomath}
in a region similar to our common selection with
the jet requirements removed. The numerator $n_{\text{data}}$ represents the number of events passing the stringent photon selection. The denominator $N^{\text{unweighted}}_{\text{fake}}$ represents the number of events passing the nonprompt-enriched photon selection mentioned in Sec~\ref{sec:event rec}. The contribution from the signal region is removed for both the numerator and denominator, and the prompt contribution in the denominator is also removed by subtracting the small prompt photon contribution from the data based on simulated samples. The factor $\epsilon_{\text{fake-fraction}}$ is the fraction of nonprompt photons in the region where the stringent photon selection is applied, which is obtained from a template fit of the photon $\sigma_{\mathrm{i}\eta\mathrm{i}\eta}$ distribution, since the variable $\sigma_{\mathrm{i}\eta\mathrm{i}\eta}$ quantifies the width of the photon electromagnetic shower in $\eta$, which is narrow for prompt and broad for nonprompt photons. The prompt-photon template is obtained from simulated $\PZ\PGg$ events and the nonprompt-photon template is obtained from a sideband method of inverting the charged hadron isolation variable of the photon in data. The event weight of an event containing a nonprompt photon can be then calculated as a function of $\pt^{\PGg}$ for photons in the barrel and endcap regions. The nonprompt-photon background estimate in the signal region is determined by these event weights and the rate of events passing the signal region selection with the stringent photon selection replaced by the nonprompt-enriched photon selection.

\section{Systematic uncertainties}
\label{sec:uncertainties}

The sources of systematic uncertainty can be divided into experimental and theoretical categories. The experimental sources include uncertainties in corrections to the simulation, the method to estimate the nonprompt-photon background contribution, and corrections for detector effects during data taking not properly accounted by the simulation in the experiment. The sources of theoretical uncertainty include the choice of the renormalization and factorization scales, and the choice of the PDFs.

The uncertainty because of renormalization and factorization scales denoted as $\mu_{\mathrm{R}}$ and $\mu_\mathrm{F}$, respectively, is evaluated for the signal and QCD-induced $\PZ\PGg$jj background. The different choices for $\mu_{\mathrm{F}}$ and $\mu_{\mathrm{R}}$ considered are these six combinations: $(\mu_{\mathrm{F}}$, $\mu_{\mathrm{R}}) = (2\mu_0$, $\mu_0)$, $(0.5\mu_0$, $\mu_0)$, $(\mu_0$, $2\mu_0)$, $(\mu_0$, $0.5\mu_0)$, $(2\mu_0$, $2\mu_0)$, and $(0.5\mu_0$, $0.5\mu_0)$, in which $\mu_0$ represents the nominal scale. These six combinations are further divided into three groups according to either $\mu_{\mathrm{R}}$ or $\mu_{\mathrm{F}}$ not equalling $\mu_0$ or both of them not equalling $\mu_0$. The difference in one group is calculated and regarded as one component of the uncertainties in $\mu_{\mathrm{R}}$ and $\mu_\mathrm{F}$. The uncertainties of these three components range from 1 to 12\% for the EW signal and from 6 to 25\% for QCD-induced $\PZ\PGg$jj background. All three components are included as independent systematic uncertainties in the ML fit introduced in Sec~\ref{sec7}. The PDF and related strong coupling $\alpS$ are evaluated using the eigenvalues of the PDF set following the NNPDF prescription~\cite{Butterworth:2015oua}. The size of this uncertainty is 1\%--3\% for both EW signal and QCD-induced $\PZ\PGg$jj background.

The uncertainties in the jet modeling, which include the JES, the jet energy resolution (JER), and the PU jet identification, are calculated in simulated events. The JES and JER are obtained by applying the corrections shifted by $\pm$1 SD. The effect of the updated corrections is propagated to all dependent variables and all selection criteria, and these effects on the yield are determined bin-by-bin for each bin of the $m_{\mathrm{jj}}$-$\abs{\Delta\eta_{\mathrm{jj}}}$ distribution. The variation, $+1$ SD or $-1$ SD, that has the larger absolute effect on the yield is assumed as the uncertainty. The size of the uncertainty varies between 1 and 92\%, depending on the $m_{\mathrm{jj}}$-$\abs{\Delta\eta_{\mathrm{jj}}}$ bin, but the larger values typically correspond to bins where the uncertainty is less important because they are applied to a smaller nominal yield. The uncertainties in PU jet identification are calculated by changing the corresponding scale factors $\pm1$ SD, depending on \pt and $\eta$. The uncertainties in jet energy corrections and PU jet working points are uncorrelated between different years, but correlated between the electron and muon categories.

The systematic uncertainty in the nonprompt-photon background estimate is the quadratic sum of three components. The first component is the uncertainty in the choice of the isolation sideband. The second component is the uncertainty in the bias in the fitting procedure, calculated by performing the procedure in simulated pseudodata and comparing the fit results with the known fractions. This component, which is larger in the endcap than in the barrel, increases with photon \pt. The third component is the uncertainty in the modeling of the prompt-photon events in the true template fit, estimated as the difference between prompt-photon events simulated from QCD-induced $\PZ\PGg$ and EW $\PZ\PGg$jj signal. The overall uncertainty in the nonprompt-photon background estimation ranges from 9 to 37\%, uncorrelated between years and channels.

The uncertainties that arise from the finite number of events in simulated samples and data control regions, referred to as statistical uncertainties, are calculated bin-by-bin based on the Poisson distribution. The statistical uncertainties typically increase with increasing $m_{\mathrm{jj}}$ and $\etajj$ and are uncorrelated across different processes and bins. The uncertainties in the efficiencies of lepton identification, trigger efficiencies, and photon identification range from 0.5 to 3.0\% depending on \pt and $\eta$ and include both statistical and systematic sources. The systematic (statistical) uncertainties are dominant in the low (high) \pt range.

All simulated samples are also affected by uncertainties associated with the ECAL timing shift, the reweighting of the PU distribution, and the integrated luminosity. The uncertainties associated with the ECAL timing shift correction factors range from 1 to 4\%, depending on the process and the $m_{\mathrm{jj}}$-$\abs{\Delta\eta_{\mathrm{jj}}}$ bin. The uncertainty from pileup reweighting is evaluated by changing the total inelastic cross section of 69.2\unit{mb}~\cite{minibais} by $\pm$5\%, which results in an uncertainty in the 1\%--10\% range. The integrated luminosity of the 2016, 2017, and 2018 data-taking periods have uncertainties in the 1.2--2.5\% range~\cite{CMS-LUM-17-003,CMS:2018elu,CMS:2019jhq}. These uncertainties are partially correlated and correspond to a total uncertainty of 1.6\%. The uncertainties of ECAL mistiming, reweighting  of the PU, and integrated luminosity uncertainties are correlated across years.

All of the above systematic uncertainties are applied in the calculation of the signal significance, measurements of the cross section, and in the search for aQGCs. They are also applied in the cross section measurement, with the exception of the theoretical uncertainties related to the normalization of the signal.

We organize the systematic uncertainties into the following groups: theoretical uncertainties including the scales $\mu_{\mathrm{R}}$ and $\mu_\mathrm{F}$, and the PDF uncertainties; corrections applied to jets including JES and JER; uncertainties in the nonprompt-photon background estimate; statistical uncertainties from simulation or data; corrections applied to electron and photon candidates; corrections applied to muon candidates; uncertainties from PU reweighting; uncertainties in the integrated luminosity; uncertainties from the L1 trigger timing shift; and uncertainties from corrections of PU jet identification (ID) working points. The remaining uncertainties, which are referred to as ``other'', include the uncertainties in the cross section estimation of diboson and {\ttbar}$\gamma$ processes that have an impact $<$0.1\%. The uncertainties in $\mu_{\mathrm{R}}$ and $\mu_\mathrm{F}$ of the QCD-induced production and in the jet energy correction are the dominant systematic uncertainties in the measurement. The impact of the uncertainty of each group on the signal strength measurement is displayed in Table~\ref{tab:sys_unc}.

\begin{table*}[htb]
\centering
\topcaption{The impact of the systematic uncertainties on the EW signal strength measurement.}
\begin{scotch}{lcc}
Systematic uncertainty& \multicolumn{2}{c}{Impact [\%] }\\
\hline
Jet energy correction                   & +7.9 & $-$6.7\\
Theoretical uncertainties               & +5.5 & $-$4.7\\
MC statistical uncertainties            & +4.7 & $-$4.5\\
PU                                      & +4.7 & $-$4.1\\
Related to \Pe, $\gamma$                & +4.5 & $-$3.6\\
PU jet ID                               & +3.7 & $-$3.4\\
ECAL timing shift at L1                 & +3.5 & $-$2.8\\
Nonprompt-$\gamma$ bkg. estimate        & +2.0 & $-$1.6\\
Related to $\mu$                        & +1.7 & $-$1.4\\
Integrated luminosity                   & +0.8 & $-$0.6\\[\cmsTabSkip]
Total systematic uncertainty            & +14  & $-$12\\
\end{scotch}
\label{tab:sys_unc}
\end{table*}

\section{Results}
\label{results}
\begin{figure*}[htbp]
   \centering
      \includegraphics[width=0.48\textwidth]{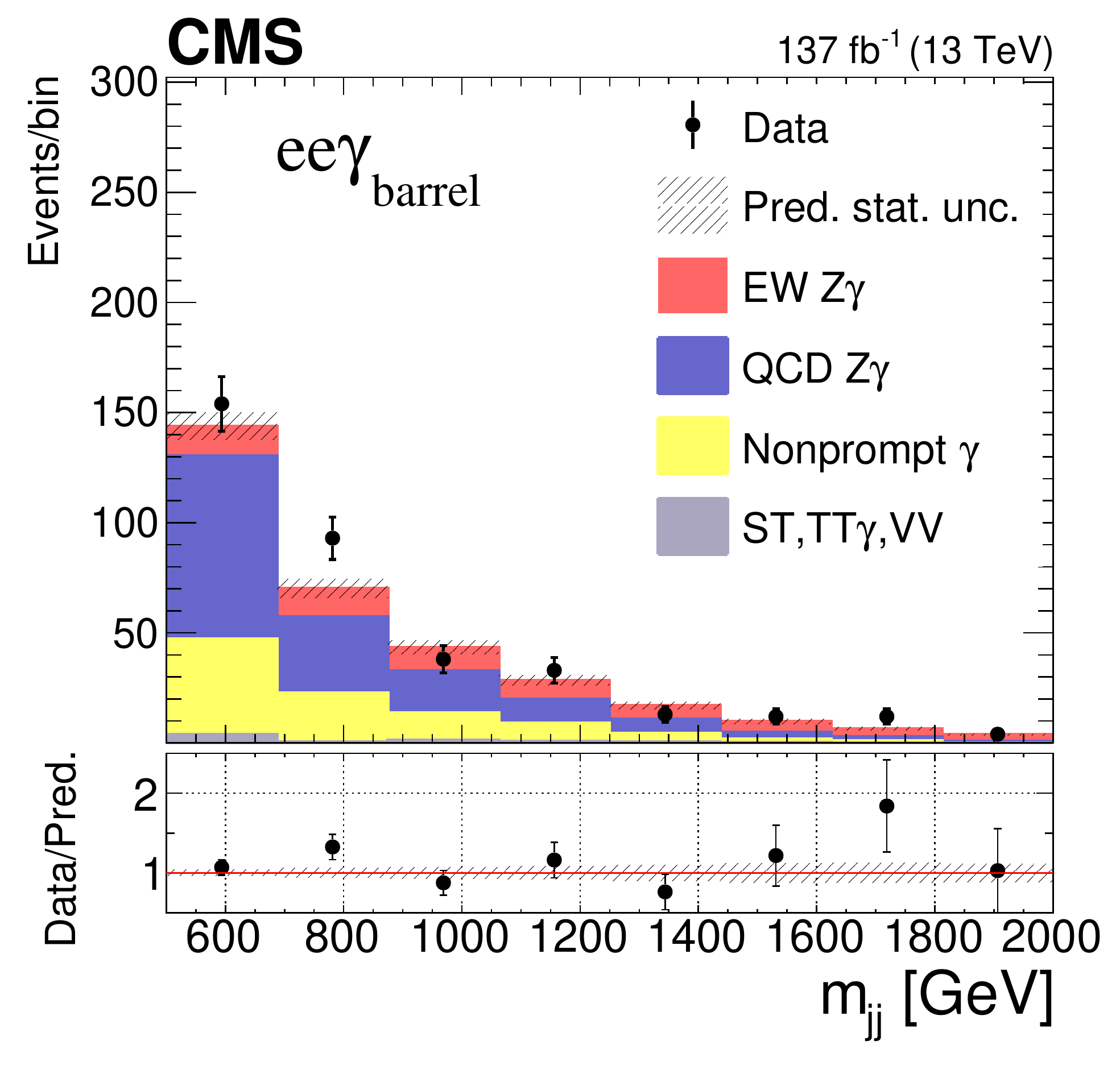}
      \includegraphics[width=0.48\textwidth]{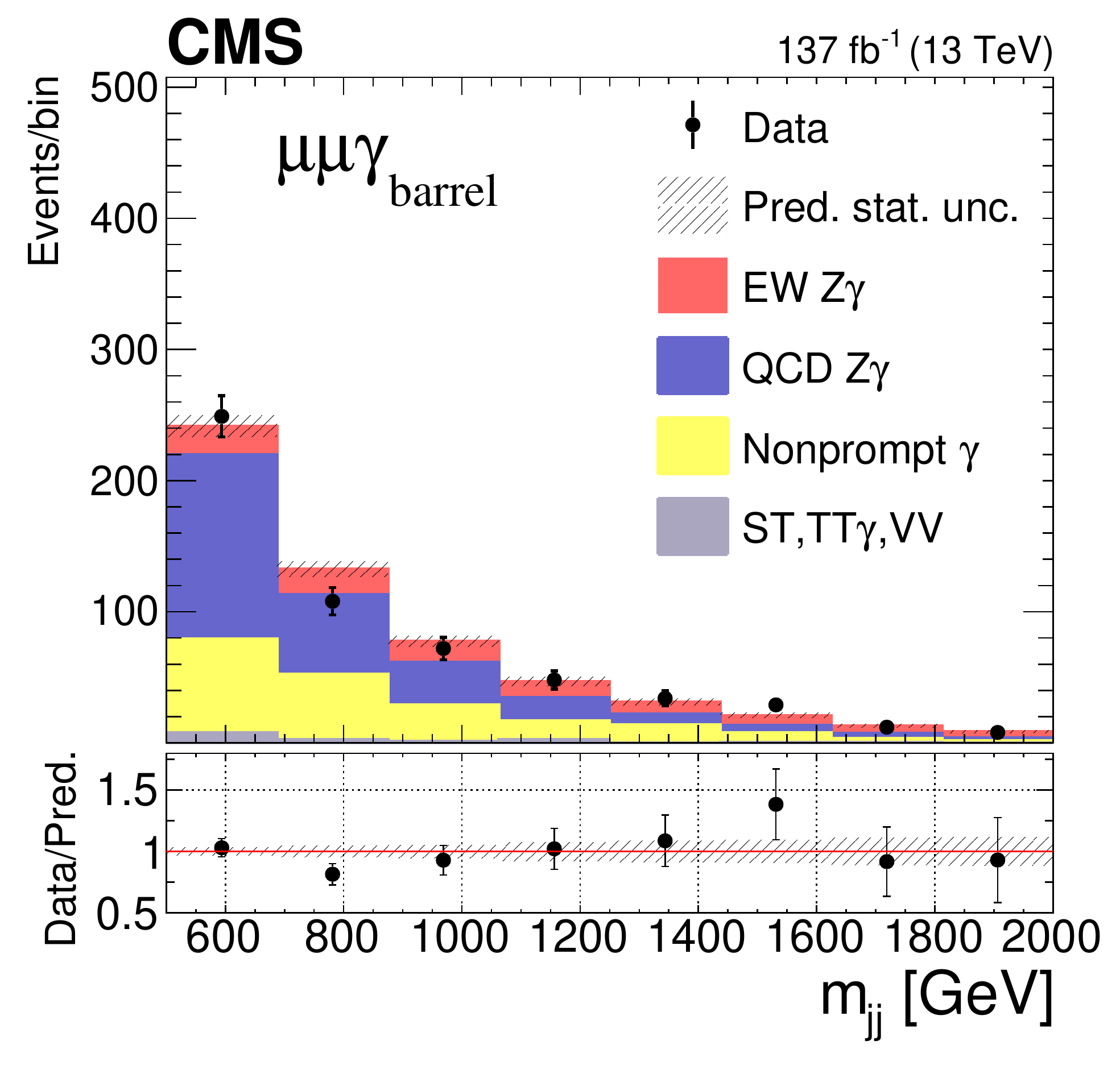}
      \caption{The pre-fit $m_{\mathrm{jj}}$ distributions in the signal region for the dilepton+$\gbarrel$ events are shown for the dielectron (left) and the dimuon (right) categories with data collected from 2016 to 2018. The data are compared to the sum of the signal and the background contributions. The black points with error bars represent the data and their statistical uncertainties, whereas the hatched bands represent the statistical uncertainty in the combined signal and background expectations. The last bin includes overflow events. The lower panel shows the ratio of the data to the expectation.}
      \label{fig:mjj_distri_b_full}
\end{figure*}

\begin{figure*}[htbp]
   \centering
      \includegraphics[width=0.48\textwidth]{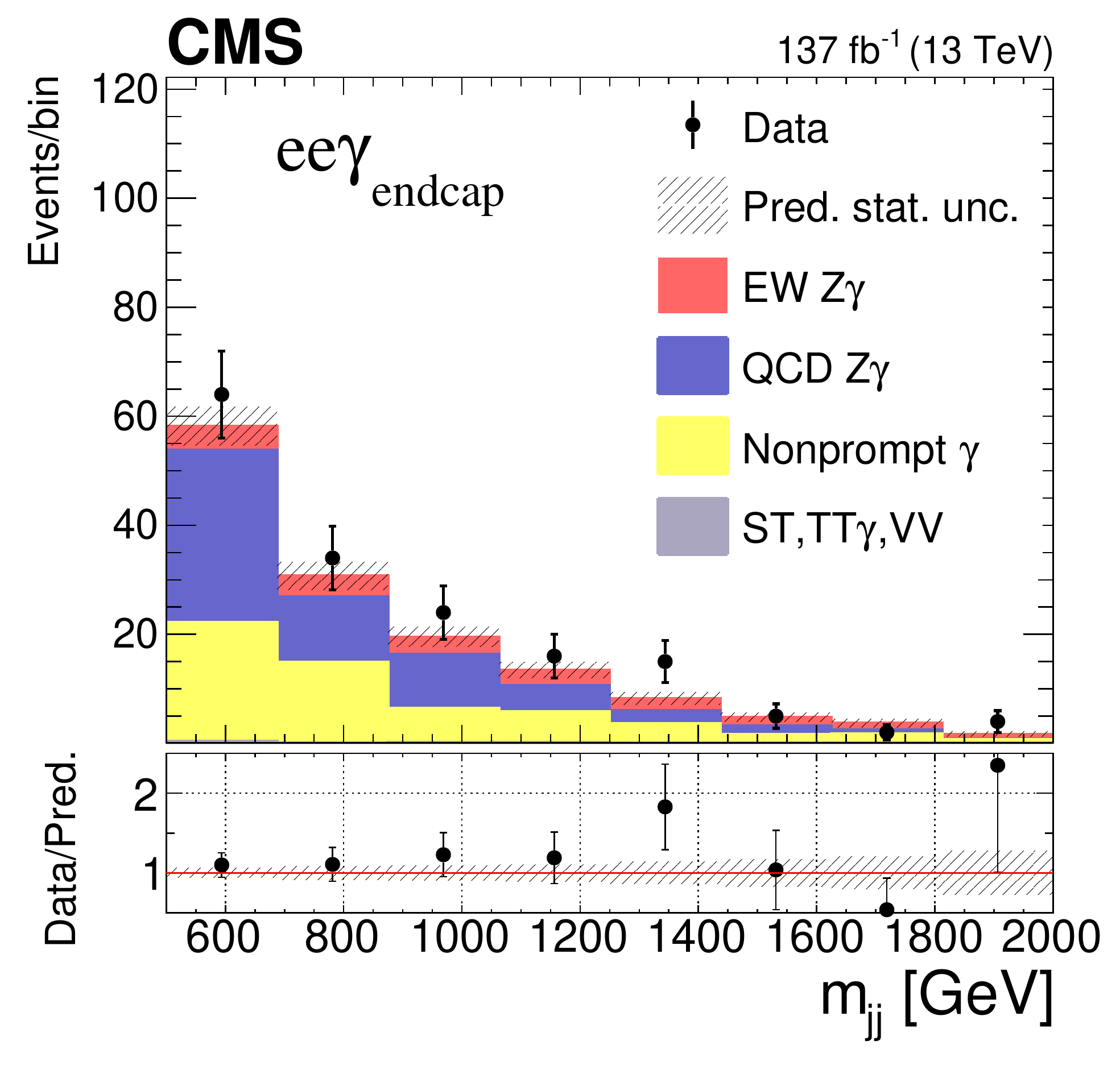}
      \includegraphics[width=0.48\textwidth]{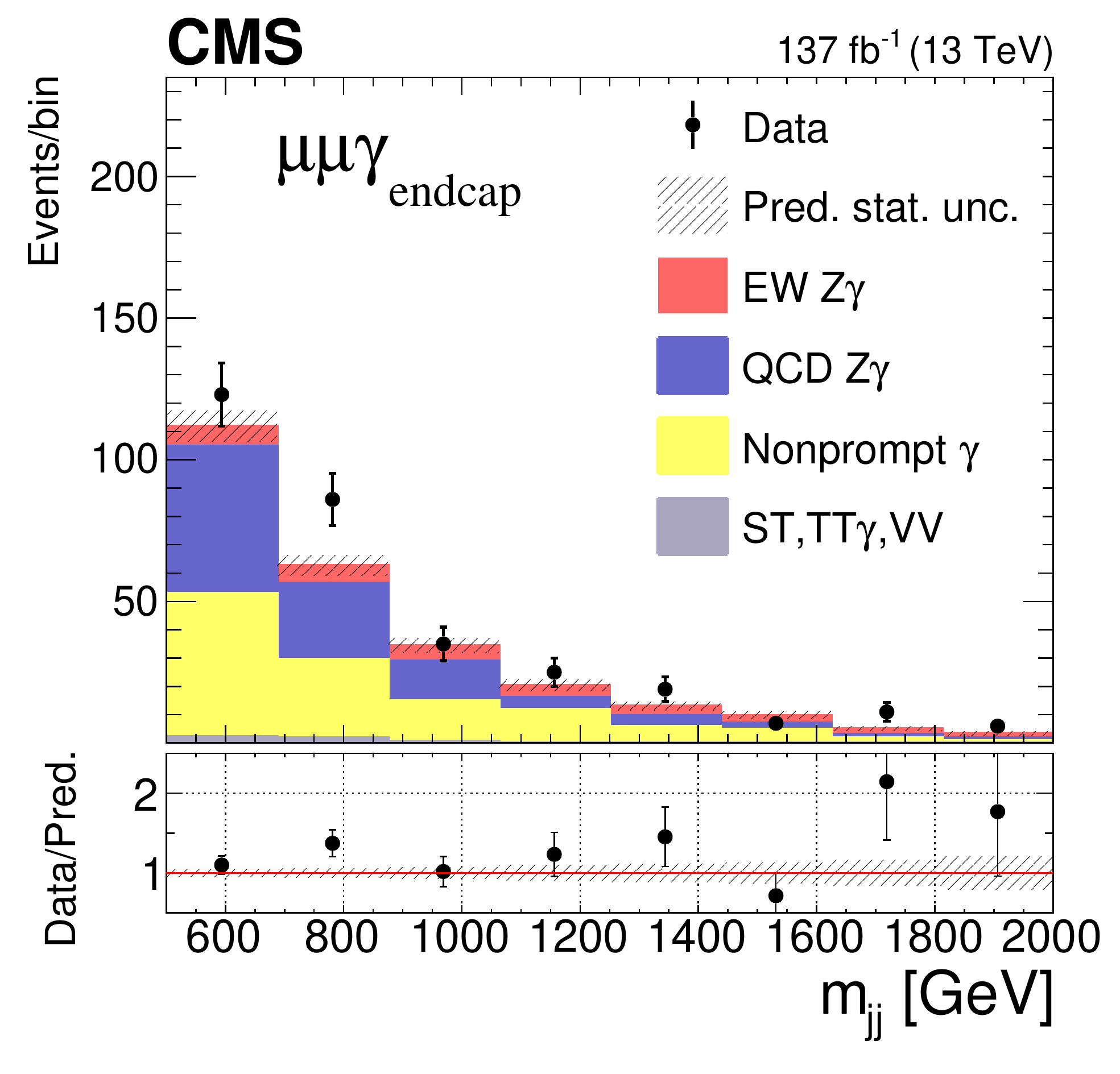}
      \caption{The pre-fit $m_{\mathrm{jj}}$ distributions in the signal region for the dilepton+$\gendcap$ events are shown for the dielectron (left) and the dimuon (right) categories with data collected from 2016 to 2018. The data are compared to the sum of the signal and the background contributions. The black points with error bars represent the data and their statistical uncertainties, whereas the hatched bands represent the statistical uncertainty in the combined signal and background expectations. The last bin includes overflow events. The lower panel shows the ratio of the data to the expectation.}
      \label{fig:mjj_distri_e_full}
\end{figure*}

The pre-fit (before the simultaneous fit described in Sec~\ref{sec7}) signal and background expected yields, as well as the observed data yields in the signal region, are shown in Figs.~\ref{fig:mjj_distri_b_full} and~\ref{fig:mjj_distri_e_full} separately for the photon in the ECAL barrel and endcaps, for both the dielectron and dimuon channels.

\subsection{Measurement of the signal significance} 
\label{sec7}
To extract as much information as possible from the data set, we perform a likelihood-based statistical analysis. An optimal binning leads to the best expected signal significance used. The likelihood function is the product of the binned signal and backgrounds probability density functions (pdf), which are expected to follow the Poisson distribution:
\begin{linenomath}
\begin{equation}\label{likelihood}
\mathcal{L} = \prod_{i}^N\mathrm{Poisson}(n_i|\mu s_i(\vec{\theta})+b_i(\vec{\theta})) p(\tilde{\vec{\theta}}|\vec{\theta}),
\end{equation}
\end{linenomath}
where $n_i$ is the number of observed events in data, $s_i$ and $b_i$ are the expected event yields for the signal and backgrounds, N represents the number of bins in the signal and control regions, and $p(\tilde{\vec{\theta}}|\vec{\theta})$ is a Gaussian constrained pdf of the effects of the systematic uncertainties $\vec{\theta}$ called nuisance parameters (NP), in which $\tilde{\vec{\theta}}$ represents the external measurements corresponding to each NP. The parameter of interest (POI) $\mu$ is the signal strength, which represents the ratio of observed to expected signal yields. The POI $\mu$ is estimated by maximizing the profile likelihood ratio defined as:
\begin{linenomath}
\begin{equation}
\label{test-stat}
\lambda(\mu)=\frac{\mathcal{L}(\mu,\hat{\hat{\vec{\theta}}})}{\mathcal{L}(\hat{\mu},\hat{\vec{\theta}})}.
\end{equation}
\end{linenomath}
The numerator of this ratio is the profile likelihood function. The quantity $\hat{\hat{\vec{\theta}}}$ denotes the value of $\vec{\theta}$
that maximizes $\mathcal{L}$ for the specified $\mu$; it is the conditional ML estimator of $\vec{\theta}$ (and thus
is a function of $\mu$). The denominator is the maximized (unconditional) likelihood function, $\hat{\mu}$ and $\hat{\vec{\theta}}$ are their ML estimators. The presence of the nuisance parameters broadens the profile likelihood as a function of $\mu$ relative to that if their values were fixed. This reflects the loss of information about $\mu$ because of the systematic uncertainties. If $\mu=0$, there are no signal events, corresponding to the background-only hypothesis. A test statistic is defined as $t_{\mu}=-2\ln\lambda(\mu)$. Higher values of $t_{\mu}$ thus correspond to increasing incompatibility between the data and the background-only hypothesis. To quantify the level of disagreement between the data and the background-only hypothesis, the $p$-value~\cite{pvalue} is computed using Eq.~(\ref{p-value}), where $t_{0}=-2\ln\mathcal{L}(0,\hat{\hat{\vec{\theta_0}}})/\mathcal{L}(\hat{\mu},\hat{\vec{\theta}})$ and $f(t_{0}|0)$ denotes the probability density function of the statistic $t_{0}$ under assumption of the background-only ($\mu=0$) hypothesis~\cite{Cowan:2010js}.
\begin{linenomath}
\begin{equation}\label{p-value}
p_{0}=\int_{t_{0,obs}}f(t_{0}|0)\,\rd t_{0}.
\end{equation}
\end{linenomath}
The $p$-value is then converted to a significance based on the area in the tail of a normal distribution. 

\begin{figure*}[ht!]
   \centering
      \includegraphics[width=0.48\textwidth]{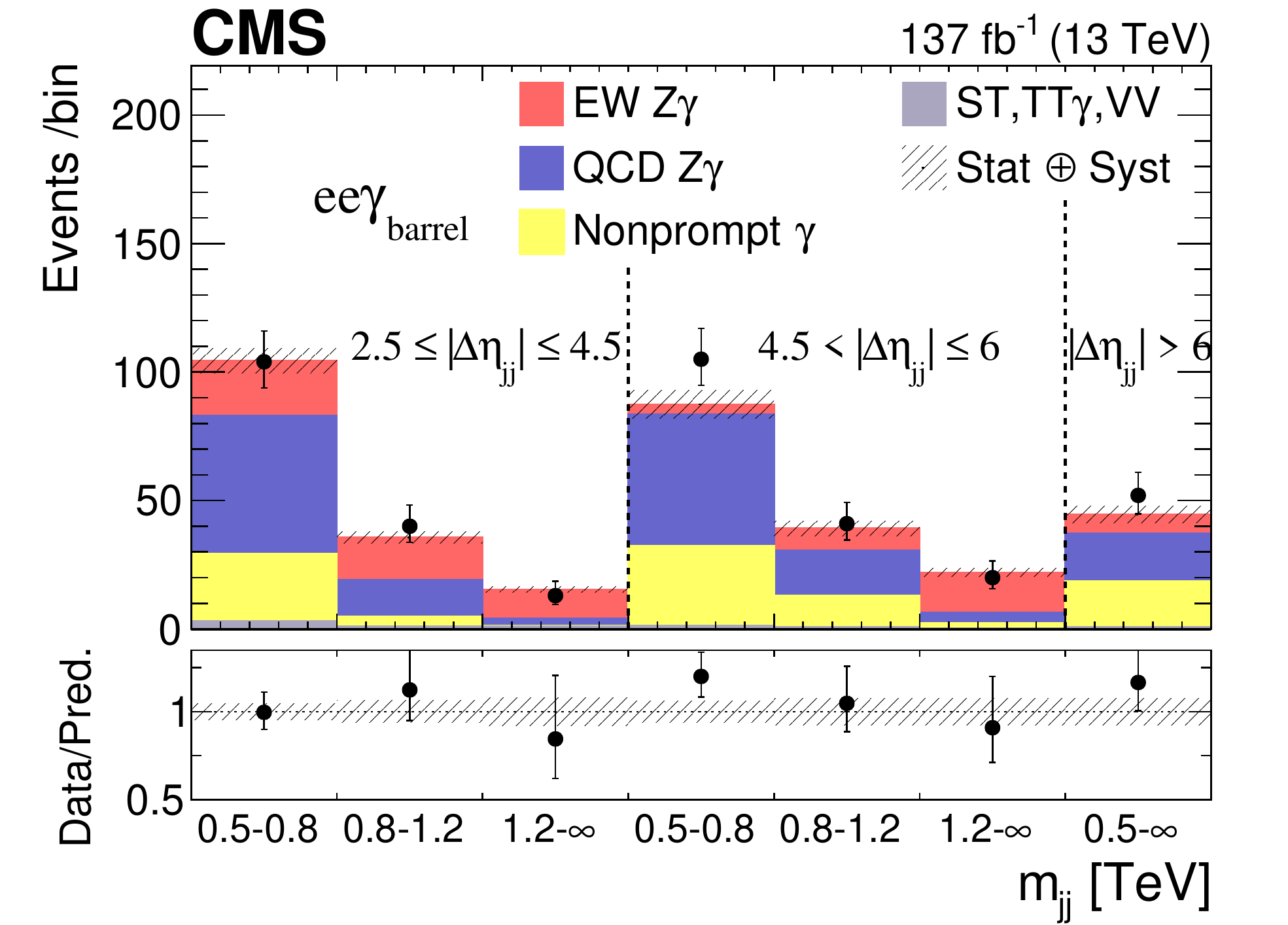}
      \includegraphics[width=0.48\textwidth]{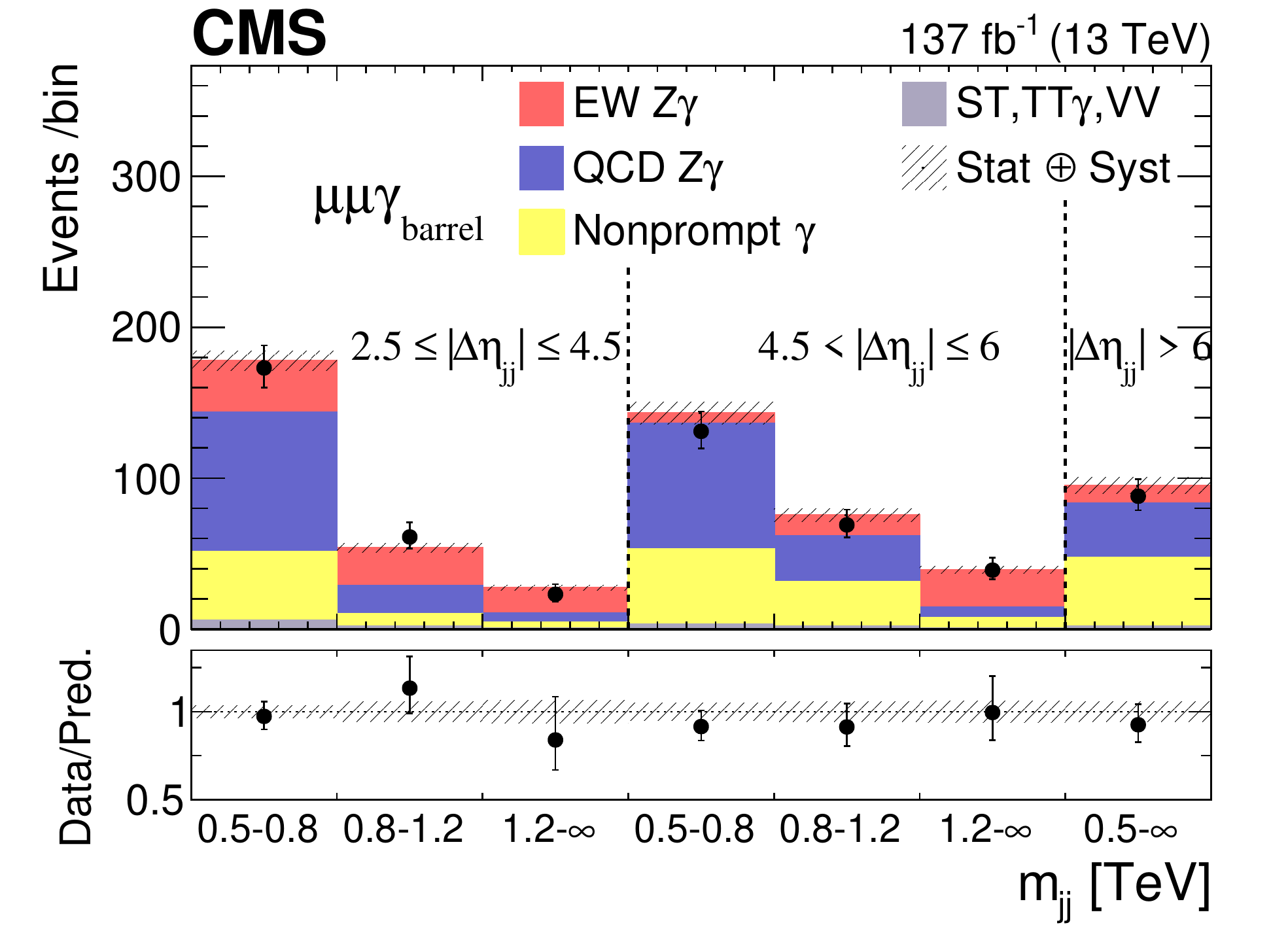}
      \caption{The post-fit 2D distributions of the dielectron (left) and dimuon (right) for the $\gbarrel$ categories in the signal region, as functions of $\mjj$ in bins of $\etajj$. The horizontal axis is split into bins of $\etajj$ of [2.5, 4.5], [4.5, 6.0], and $>$6.0. The data are compared to the signal and background in the predictions. The black points with error bars represent the data and their statistical uncertainties, whereas the hatched bands represent the total uncertainties of the predictions.}
      \label{fig:prefit_b}
\end{figure*}

\begin{figure*}[ht!]
   \centering
      \includegraphics[width=0.48\textwidth]{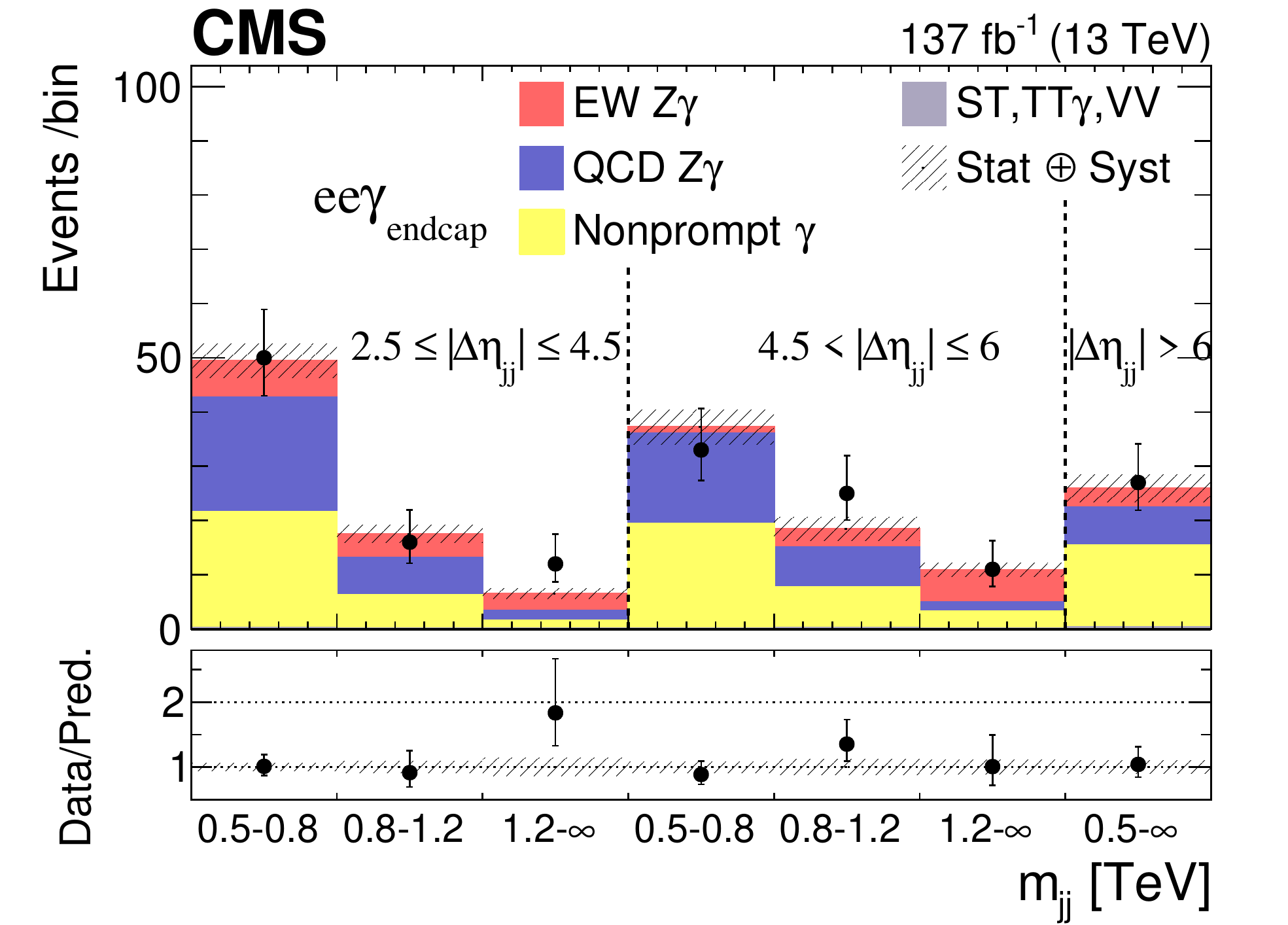}
      \includegraphics[width=0.48\textwidth]{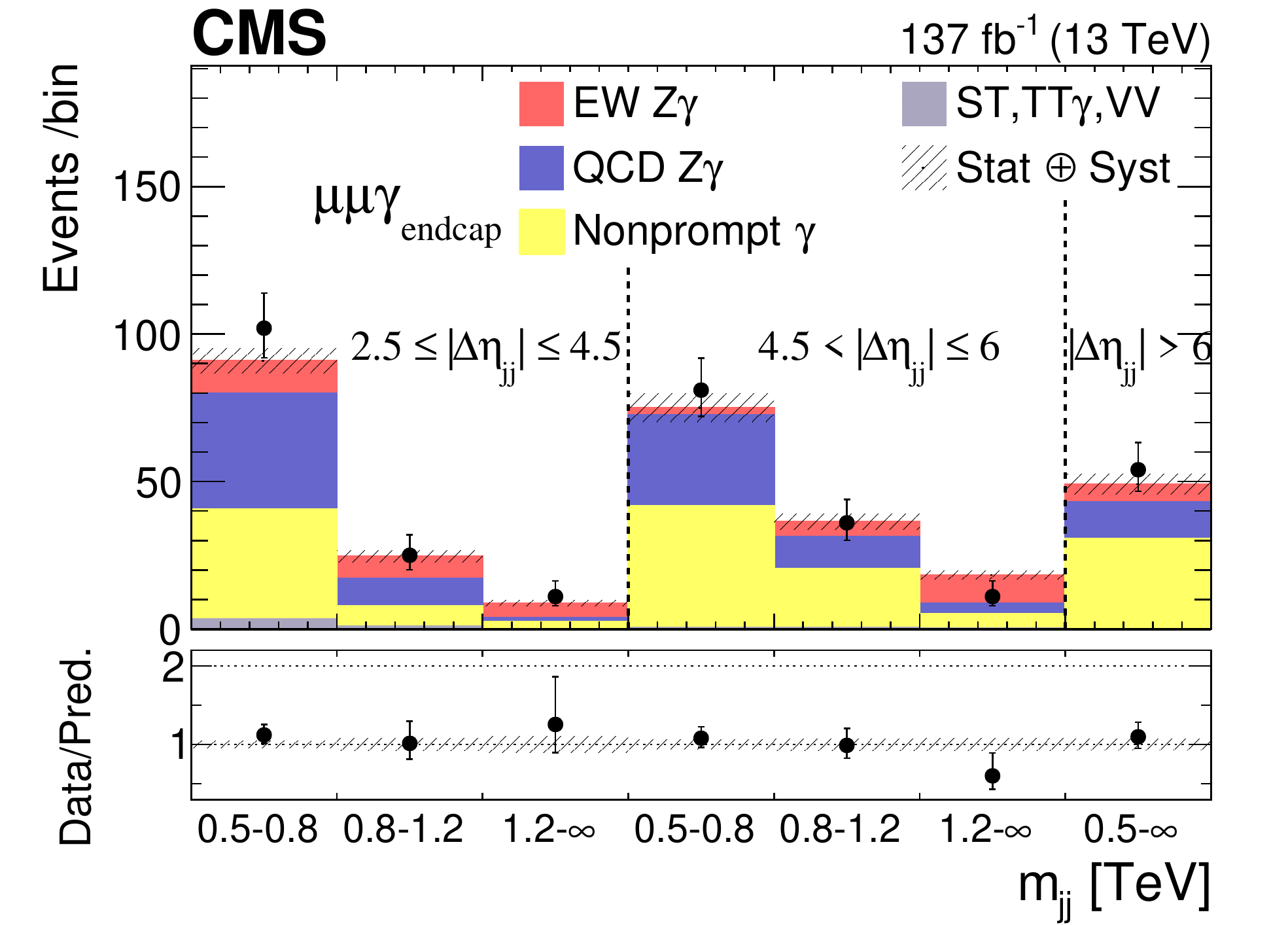}
      \caption{The post-fit 2D distributions of the dielectron (left) and dimuon (right) for the $\gendcap$ categories in the signal region, as functions of $\mjj$ in bins of $\etajj$. The horizontal axis is split into bins of $\etajj$ of [2.5, 4.5], [4.5, 6.0], and $>$6.0. The data are compared to the signal and background in the predictions. The black points with error bars represent the data and their statistical uncertainties, whereas the hatched bands represent the total uncertainties of the predictions.}
      \label{fig:prefit_e}
\end{figure*}

\begin{figure*}[ht!]
   \centering
      \includegraphics[width=0.48\textwidth]{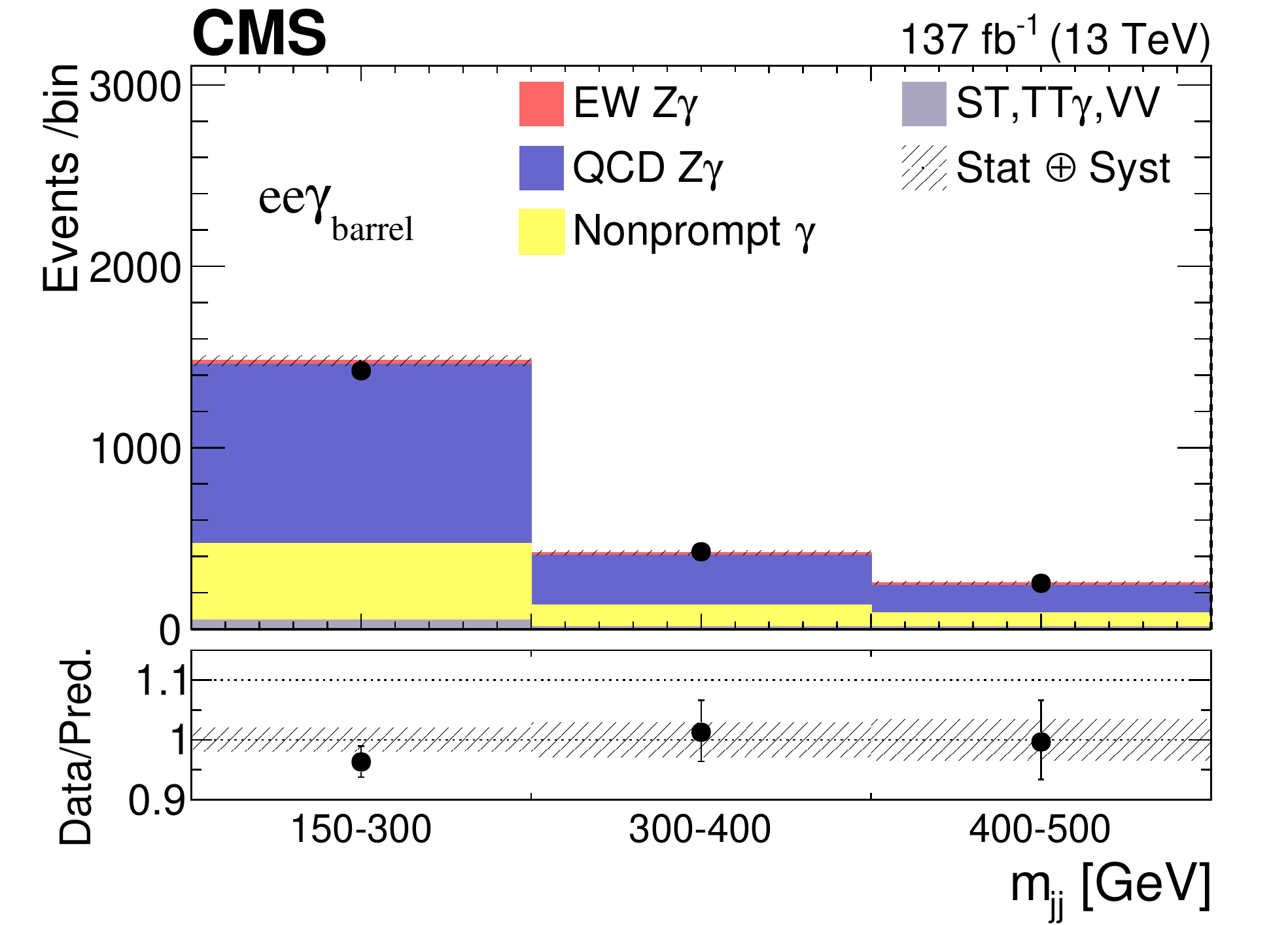}
      \includegraphics[width=0.48\textwidth]{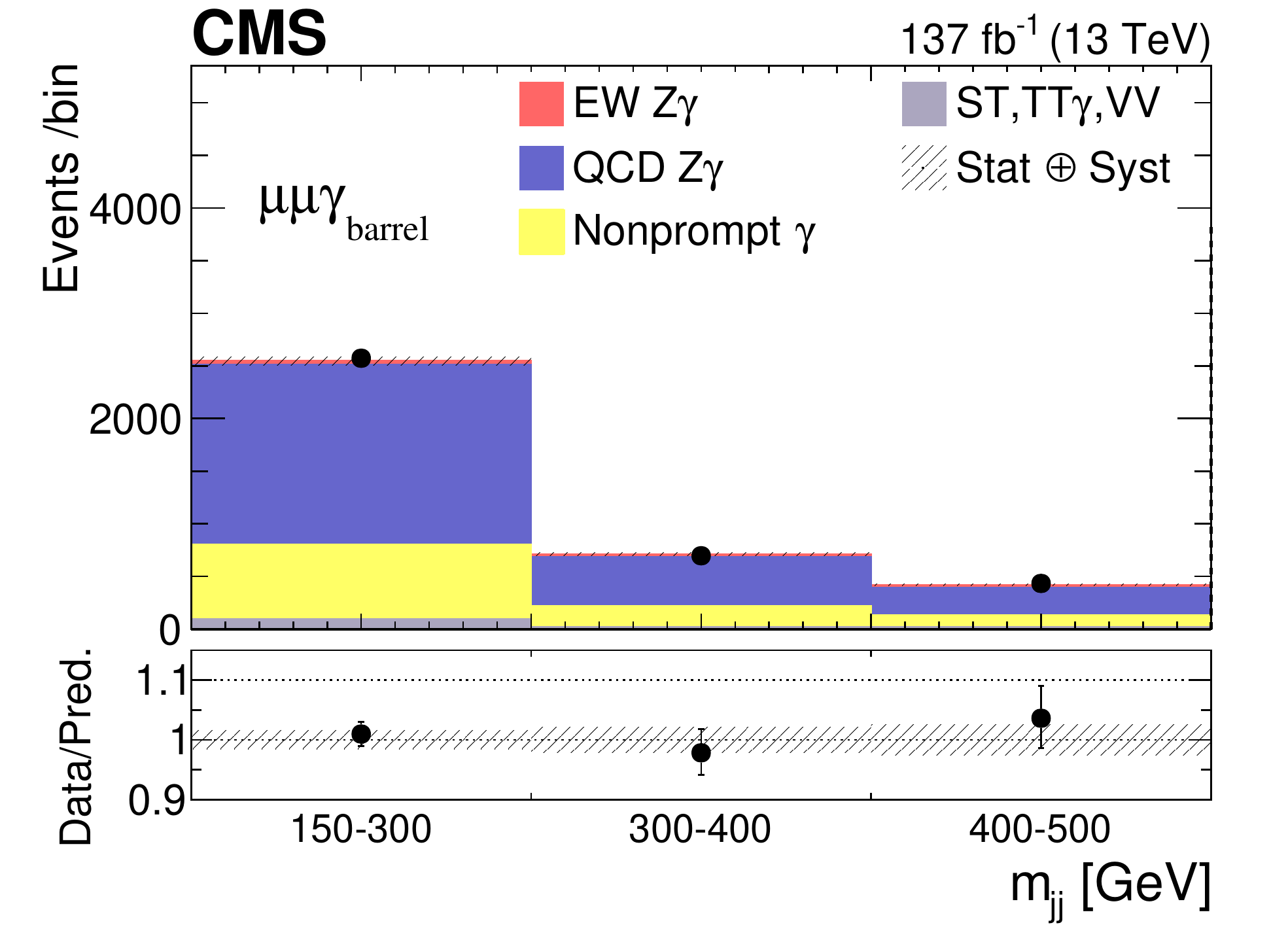}
      \caption{The postfit distributions in the control region for the dielectron (left) and dimuon (right) for the $\gbarrel$ categories as a function of $\mjj$. The horizontal axis is split into bins of $m_{\mathrm{jj}}$ of [150, 300], [300, 400], and [400,500]. The black points with error bars represent the data and their statistical uncertainties, whereas the hatched bands represent the total uncertainties of the predictions.}
      \label{fig:prefit_CR_b}
\end{figure*}

\begin{figure*}[ht!]
   \centering
      \includegraphics[width=0.48\textwidth]{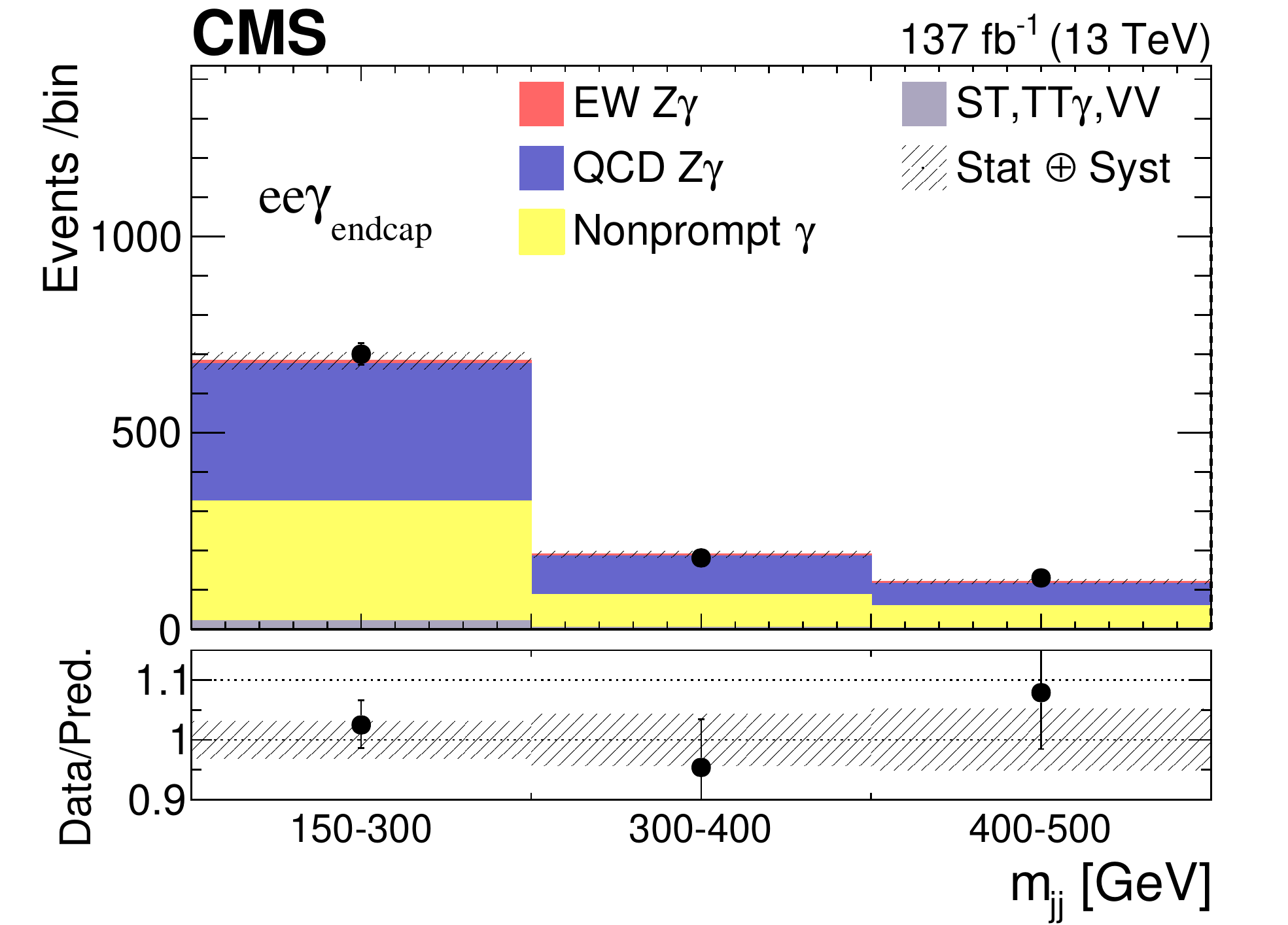}
      \includegraphics[width=0.48\textwidth]{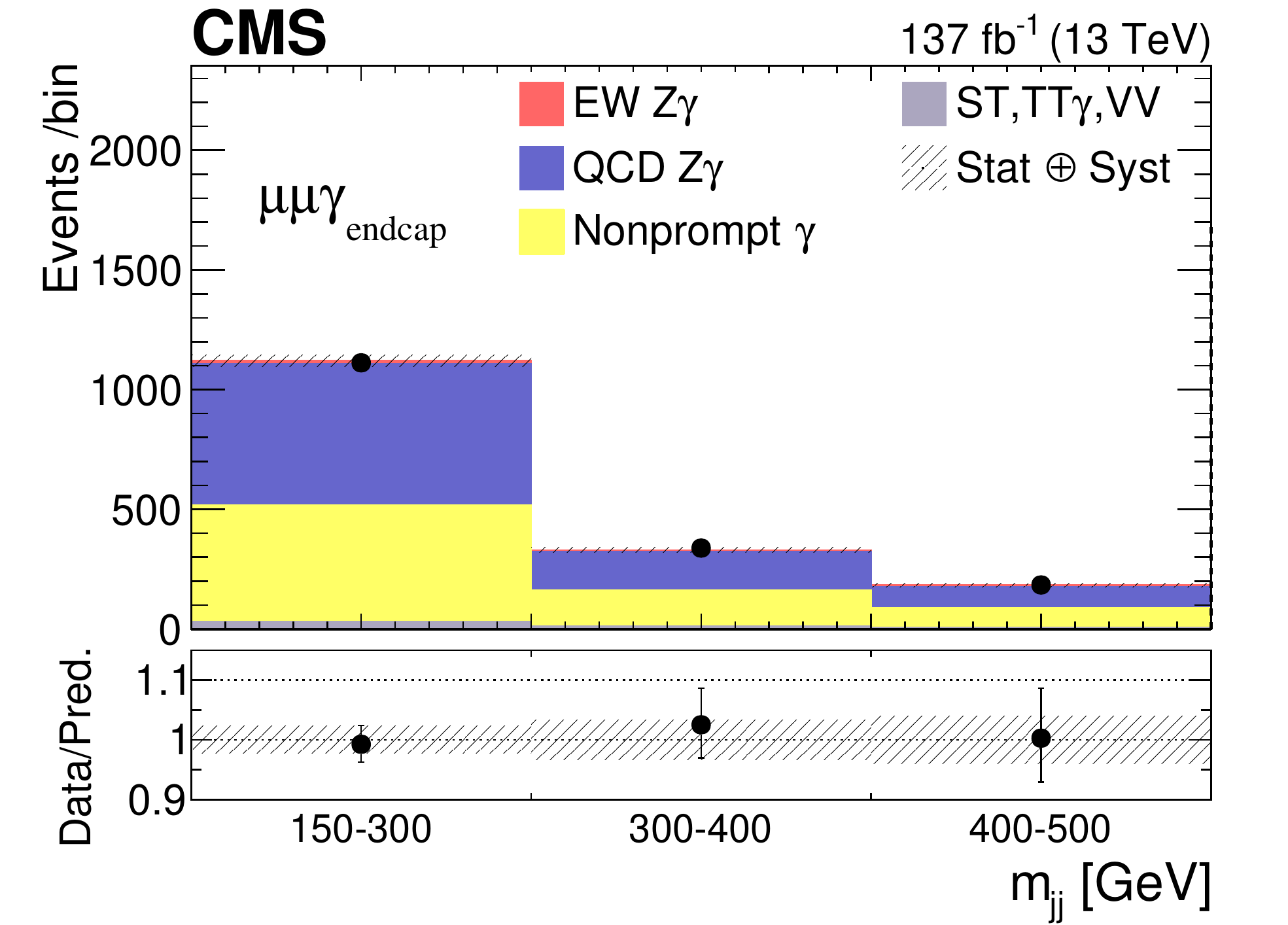}
      \caption{The postfit distributions in the control region for the dielectron (left) and dimuon (right) for the $\gendcap$ categories as a function of $\mjj$. The horizontal axis is split into bins of $m_{\mathrm{jj}}$ of [150, 300], [300, 400], and [400,500]. The black points with error bars represent the data and their statistical uncertainties, whereas the hatched bands represent the total uncertainties of the predictions.}
      \label{fig:prefit_CR_e}
\end{figure*}

The control and signal regions are each divided separately for photons in the ECAL barrel/endcaps and for the dielectron and dimuon channels. The signal region is divided further into bins in $m_{\mathrm{jj}}$ and $\etajj$, as shown in Figs.~\ref{fig:prefit_b} and~\ref{fig:prefit_e}. The control region is divided into bins in $m_{\mathrm{jj}}$, as shown in Figs.~\ref{fig:prefit_CR_b} and \ref{fig:prefit_CR_e}. The postfit yields for every process and for data are listed in Table~\ref{tab:yields_SR}.
\begin{table*}[htb]
\centering
\topcaption{Post-fit yields of predicted signal and background with total uncertainties, and observed event counts after the selection in the EW signal region. The $\gbarrel$ and $\gendcap$ columns represent events with photons in the ECAL barrel and endcaps, respectively.}
\centering
  \begin{scotch}{ccccc}
  Process & $\mu\mu\gbarrel$ & $\mu\mu\gendcap$ & ee$\gbarrel$  & ee$\gendcap$\\\hline
ST                          & 0.7 $\pm$ 0.4     & 0.2 $\pm$ 0.2    & 0.6 $\pm$ 0.3      & 0.2 $\pm$ 0.2 \\
TT$\gamma$                  & 8.8 $\pm$ 1.3     & 2.1 $\pm$ 0.5    & 3.4 $\pm$ 0.6      & 0.2 $\pm$ 0.2 \\
$\PV\PV$                    & 6.0 $\pm$ 1.9     & 3.2 $\pm$ 1.2    & 4.1 $\pm$ 1.3      & 0.8 $\pm$ 0.3 \\
Nonprompt photon            & 189 $\pm$ 9.2     & 143 $\pm$ 6.9    & 93.6 $\pm$ 6.5     & 74.3 $\pm$ 5.0\\
QCD $\PZ\PGg$               & 274 $\pm$ 10      & 108 $\pm$ 5.6    & 162 $\pm$ 7.4      & 62.4 $\pm$ 3.9\\
EW $\PZ\PGg$                & 133 $\pm$ 4.7     & 46.5 $\pm$ 1.7   & 84.5 $\pm$ 3.1     & 28.2 $\pm$ 1.1\\[\cmsTabSkip]
Predicted yields            & 612 $\pm$ 13      & 303 $\pm$ 8      & 349 $\pm$ 9        & 166 $\pm$ 6 \\
Data                        & 584               & 320              & 375                & 174\\
  \end{scotch}
  \label{tab:yields_SR}
\end{table*}

The main contributions to the significance are from the bins in the signal region with an excess of signal relative to background events, such as high $\mjj$ bins in each channel. 

The observed (expected) significance for EW signal obtained from a simultaneous fit of all bins in the signal region and the control region to the data (Asimov data set~\cite{Cowan:2010js}) is 9.4 (8.5 SD). 

\subsection{Fiducial cross section}\label{sec:fid}
Fiducial cross sections are measured in the fiducial region, which is designed to mirror the signal region as closely as possible, as shown in Table~\ref{tab:selections}. The fiducial cross section is extracted using the same binning of $m_{\mathrm{jj}}$ and $\etajj$ as used in the signal significance measurement, and with the same simultaneous fit in all regions and channels, with the following exception: the events that pass the EW signal selection but fail the fiducial selection are regarded as a background. We define the fiducial cross section as:
\begin{linenomath}
\begin{equation}
\sigma^{\text{fid}} = \sigma^{\mathrm{g}} \hat{\mu} \mathrm{a}^{\mathrm{gf}},
\label{equ:equ_fid_EW}
\end{equation}
\end{linenomath}
where $\sigma^\mathrm{g}$ is the cross section for the generated signal events, $\hat{\mu}$ is the signal strength parameter, and $\mathrm{a}^\mathrm{gf}$ is the acceptance for the events generated in the fiducial region and evaluated through simulation. The theoretical fiducial cross section for the EW $\PZ\PGg$ signal at LO accuracy is $4.34\pm 0.26\,\text{(scale)}\pm 0.06\,\mathrm{(PDF)\unit{fb}}$. The best fit value for the EW $\PZ\PGg$ signal strength and the measured fiducial cross section are 
\begin{linenomath}
\begin{subequations}
\begin{align}
   \begin{split}
\mu_{\mathrm{EW}} &= 1.20^{+0.12}_{-0.12}\stat\, ^{+0.14}_{-0.12}\syst\\ &= 1.20^{+0.18}_{-0.17},
   \end{split}\\
\begin{split}
\sigma^{\text{fid}}_{\mathrm{EW}} &= 5.21\pm0.52\stat\pm 0.56\syst\unit{fb}\\ &= 5.21\pm 0.76\unit{fb}.\label{method1}
\end{split}
\end{align}
\end{subequations}
\end{linenomath}
A combined EW+QCD $\PZ\PGg$jj cross section is also measured in the same fiducial region using the same procedure, except that the control region is excluded. In this measurement, both the EW and QCD contributions are considered signal. The combined $\PZ\PGg$jj cross section is defined as
\begin{linenomath}
\begin{align}\label{fidall}
\sigma^{\text{fid}} = \hat{\mu} ( \sigma_{\mathrm{EW}}^{\mathrm{g}} \mathrm{a}_{\mathrm{EW}}^{\mathrm{gf}} + \sigma_{\mathrm{QCD}}^{\mathrm{g}} \mathrm{a}_{\mathrm{QCD}}^{\mathrm{gf}} ).
\end{align}
\end{linenomath}
The theoretical fiducial cross section for QCD $\PZ\PGg$jj production is $8.93\pm 1.70\,\text{(scale)}\pm 0.08\,\mathrm{(PDF)\unit{fb}}$. The expected fiducial cross section for the combined QCD and EW $\PZ\PGg$jj production is $13.3\pm 1.72\,\text{(scale)}\pm 0.10\,\mathrm{(PDF)\unit{fb}}$. The best fit value for the combined $\PZ\PGg$jj signal strength and the measured cross section are
\begin{linenomath}
\begin{subequations}
\begin{align}
   \begin{split}
\mu_{\mathrm{EW+QCD}} &= 1.11^{+0.06}_{-0.06}\stat\, ^{+0.10}_{-0.09}\syst \\&= 1.11^{+0.12}_{-0.11}, 
   \end{split}\\
   \begin{split}
\sigma^{\text{fid}}_{\mathrm{EW+QCD}} &= 14.7 \pm 0.80\stat \pm 1.26\syst \unit{fb}\\&=14.7 \pm 1.53\,\unit{fb}.\label{fidall2}
   \end{split}
\end{align}
\end{subequations}
\end{linenomath}
\subsection{Unfolded differential cross section distribution}
Unfolding is used to correct measured detector-level distributions to the particle-level. It accounts for the limited acceptance and efficiencies of the detector, as well as for bin-to-bin migration between the measured and corrected distribution arising from detector resolution. Simulated EW samples from MC event generators are used to perform the unfolding. Distributions obtained from the generated events correspond to the particle level (which will be referred to as ``gen''). The same distributions obtained using simulated events correspond to detector level (which will be referred to as ``reco''). In the well-defined fiducial phase space, the events in ``reco'' and ``gen'' follow the relation:
\begin{linenomath}
\begin{subequations}
\begin{align}
& y_{i}^{\text{reco}}=\sum_{j}R_{ij} x_{j}^{\text{gen}}+b_i,\label{unfolding_equ}\\
& R_{ij}=P(\text{observed in bin } i|\text{ generated in bin } j),\label{response}
\end{align}
\end{subequations}
\end{linenomath}
in which $y_{i}^{\text{reco}}$ represents observed events in data in the reconstructed bin i, $x_{j}^{\text{gen}}$ represents the events from simulation in the bin j at the particle level, and $b_i$ represents the background from simulation in the reconstructed bin i. The number of events $\sum_{j}x_{j}^{\text{gen}}$ obeys the distribution $\vec{x}\sim\mathrm{Poisson}(\vec{\lambda})$, where $\vec{\lambda}$ represents the bin means of the distributions at the particle level. So the observed number of events $\sum_{i}y_{i}^{\text{reco}}$ is Poisson distributed, $\vec{y}\sim\mathrm{Poisson}(\boldsymbol{R}\vec{\lambda}+\vec{b})$, where the element of the response matrix $\boldsymbol{R}$ is given in the Eq.~(\ref{response}). The formal solution to Eq.~(\ref{unfolding_equ}) can be written as $x_{j}^{\text{gen}}=R_{ji}^{-1}(y_{i}^{\text{reco}}-b_i)$. The estimate of the $x_{j}^{\text{gen}}$ can also be derived from the principle of ML; the corresponding likelihood is in Eq.~(\ref{unfold_ML}). 
\ifthenelse{\boolean{cms@external}}{
\begin{multline}
\label{unfold_ML}
\mathcal{L}(\vec{\mu};\vec{\theta})\\=\prod_{i}\mathrm{Poisson}(y_i|\sum_{j}R_{ij}(\vec{\theta})\mu_{j}s_j(\vec{\theta})+b_i(\vec{\theta})\,) p(\tilde{\vec{\theta}}|\vec{\theta}).
\end{multline}
}{
   \begin{equation}
      \label{unfold_ML}
      \mathcal{L}(\vec{\mu};\vec{\theta})=\prod_{i}\mathrm{Poisson}(y_i|\sum_{j}R_{ij}(\vec{\theta})\mu_{j}s_j(\vec{\theta})+b_i(\vec{\theta})\,) p(\tilde{\vec{\theta}}|\vec{\theta}).
      \end{equation}
}
This solution is based on the assumption that $\boldsymbol{R}$ is nonsingular and insensitive to small perturbations. If that is not the case, the problem is ill-posed and $\boldsymbol{R}$ is ill-conditioned, which means that a regularization~\cite{doi:10.1137/1.9780898718836.ch4} method is needed. The sensitivity to fluctuations associated with the ML solution can be quantified by the condition number of $\boldsymbol{R}$, which is a measure of how ill-conditioned the problem is. In general, if the condition number $c(\boldsymbol{R})$ is small (less than 10), then the problem can most likely be solved using the unregularized ML estimate. In this paper, an appropriate choice of binning is made to guarantee that the response matrix $\boldsymbol{R}$ is nonsingular and $c(\boldsymbol{R})<10$. The regularization parts thus are not considered in this unfolded differential cross section measurement.

\begin{figure*}[htbp!]
   \begin{center}
   \includegraphics[width=0.45\textwidth]{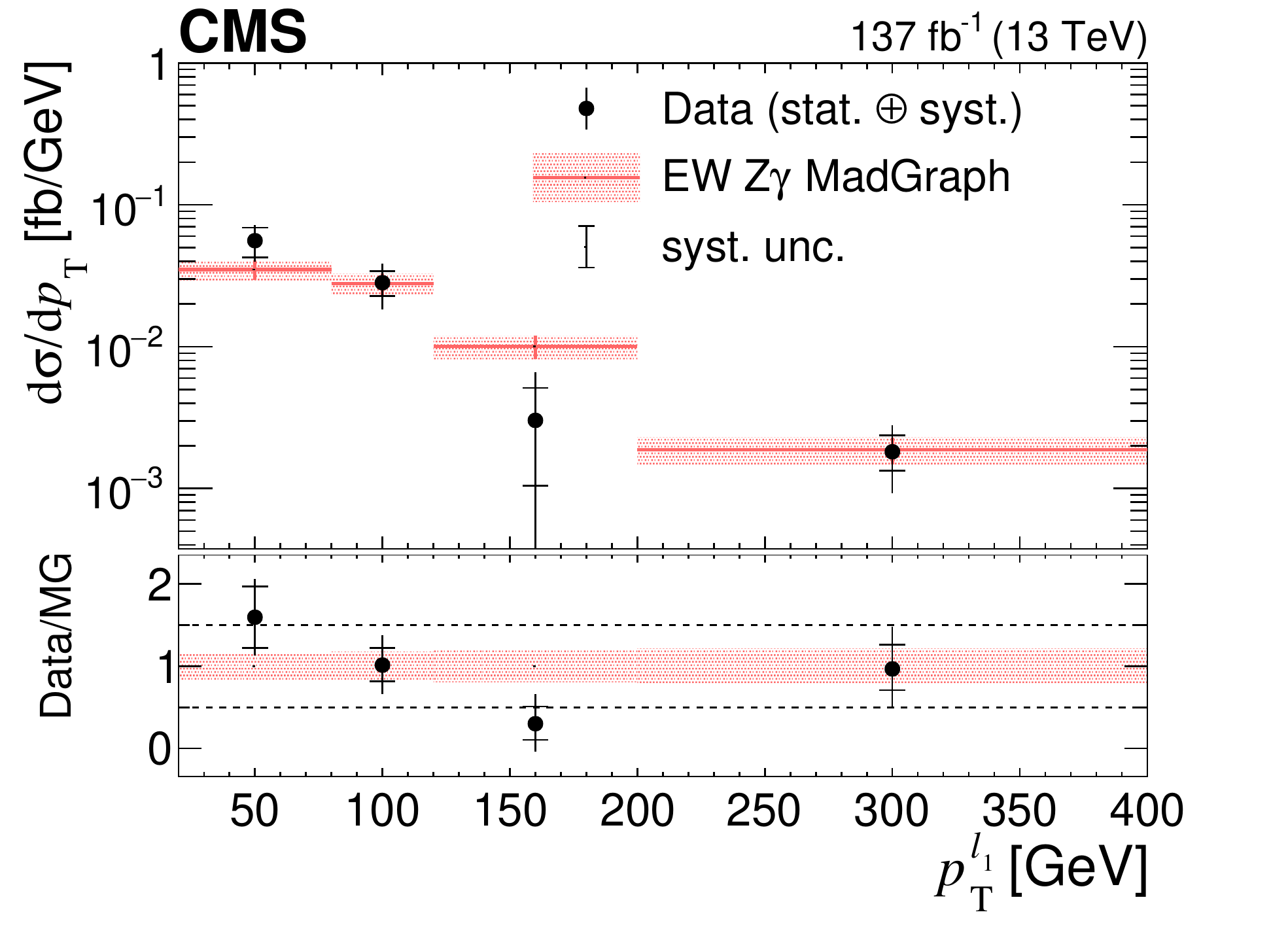}
   \includegraphics[width=0.45\textwidth]{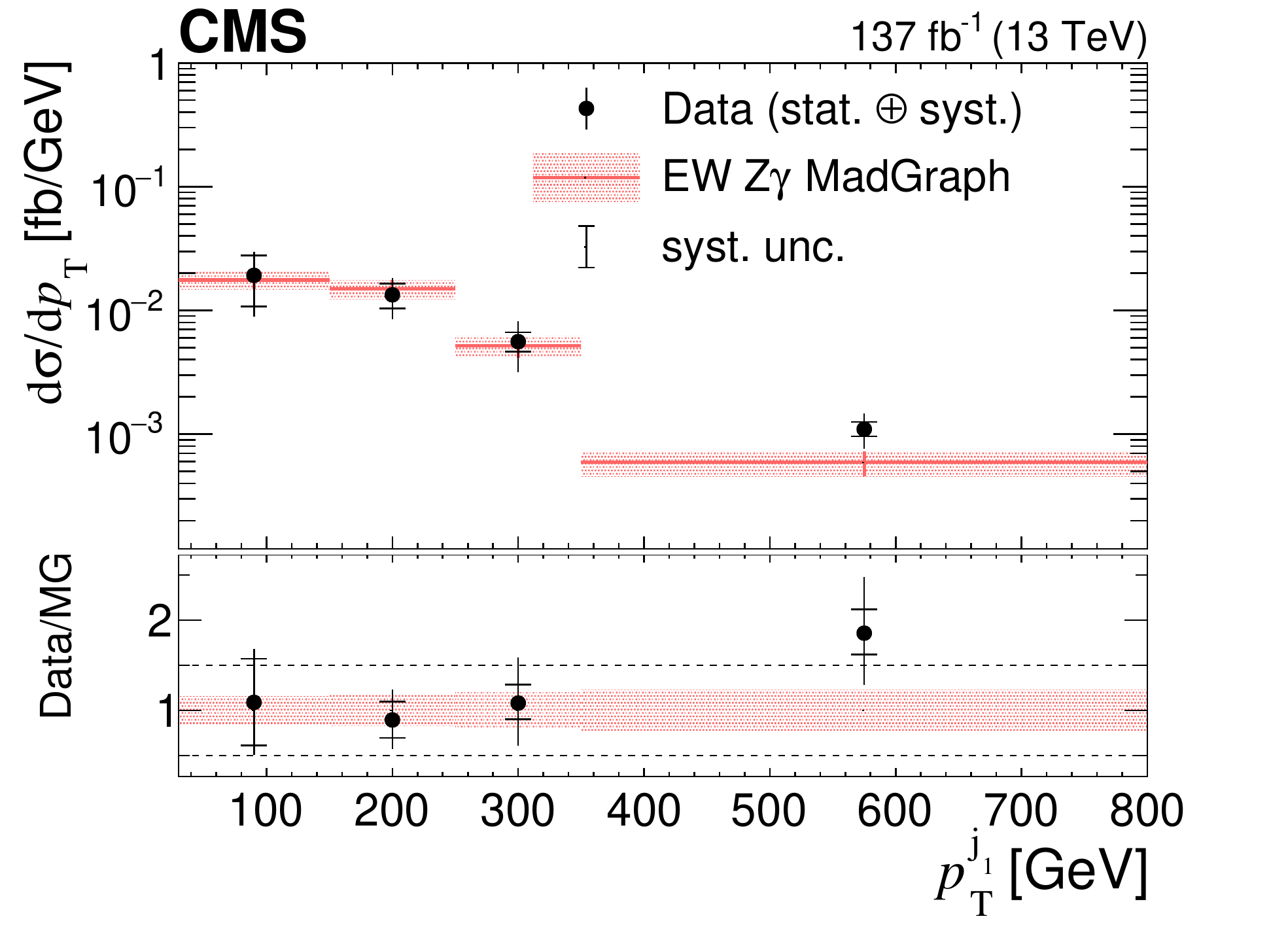}\\
   \includegraphics[width=0.45\textwidth]{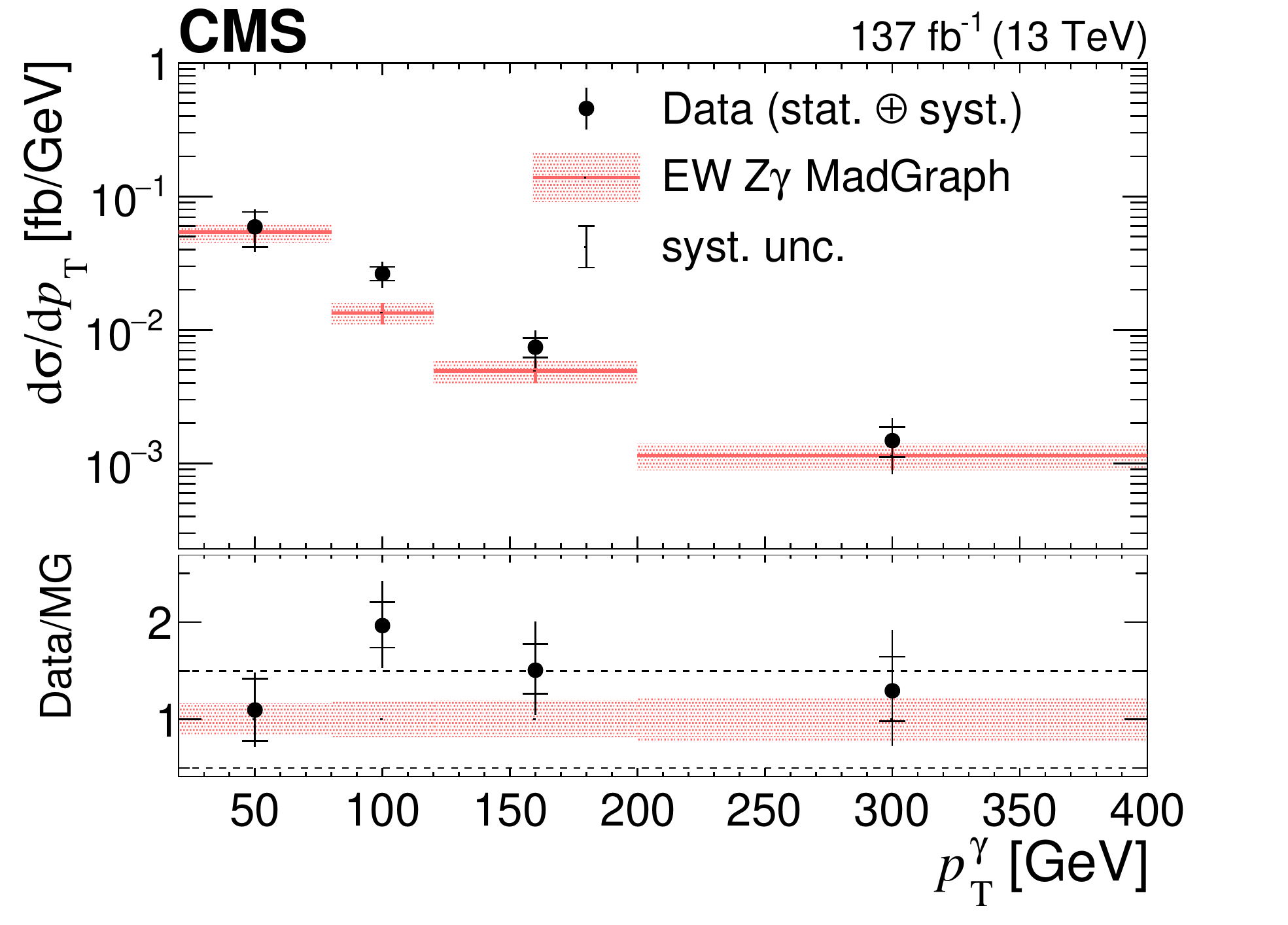}
    \includegraphics[width=0.45\textwidth]{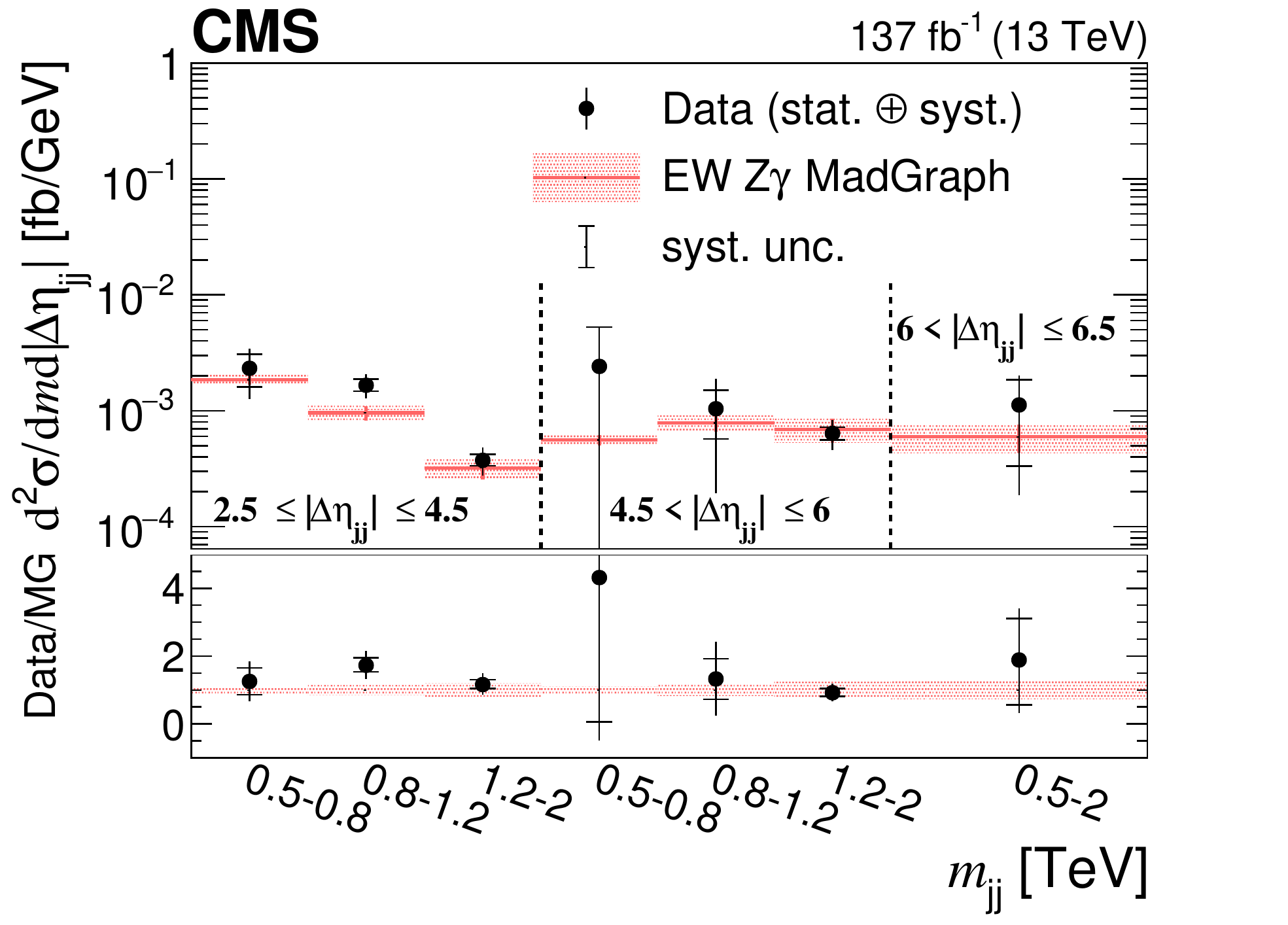}\\
   \caption{Unfolded differential cross section as a function of the leading lepton \pt, leading jet \pt, leading photon \pt, and $m_{\mathrm{jj}}$-$\abs{\Delta\eta_{\mathrm{jj}}}$ for EW $\PZ\PGg$jj. The black points with error bars represent the data and their statistical uncertainties, whereas the red bands represent the total theoretical uncertainties from the MG5 simulation. The last bin includes overflow events.}
   \label{fig:unfold_results}
   \end{center}
\end{figure*}

The unfolded differential cross section is measured in the same way as the fiducial cross section, with a simultaneous fit to data in the control region and signal region in bins of the single variables leading photon \pt, leading lepton \pt, and leading jet \pt, and in bins of two variables $m_{\mathrm{jj}}$ and $\abs{\Delta\eta_{\mathrm{jj}}}$. The off-diagonal elements of the response matrix correspond to 1\% for leading photon \pt, 1\% for leading lepton \pt, 3\% for leading jet \pt, and around 5\% for $m_{\mathrm{jj}}$ and $\abs{\Delta\eta_{\mathrm{jj}}}$. All systematic uncertainties discussed in Sec~\ref{sec:uncertainties} are also considered for the corresponding process, especially for the evaluation of $R_{ij}$. The signal strengths for each bin of the EW $\PZ\PGg$jj unfolded distribution are listed in Tables~\ref{tab:unfolding_results_1} and~\ref{tab:unfolding_results} and the corresponding differential cross section distributions are shown in Fig.~\ref{fig:unfold_results}.

\begin{table*}[hbtp]
\centering
\topcaption{The signal strengths and differential cross sections from SM expectation and fit calculated as part of the unfolding of $\pt^{\PGg}$, $\pt^{\mathrm{j}_1}$, and $\pt^{\ell_1}$ observables for EW $\PZ\PGg$jj. The last bin includes overflow events.}
\cmsTable{
\begin{scotch}{ c l c c c  }
Variables & Bin [\GeVns{}] & $\mu\pm\Delta\mu$ & Predicted $\rd\sigma/\rd\pt$ [fb/\GeVns{}] & Observed $\rd\sigma/\rd\pt$ [fb/\GeVns{}]\\
\hline \\[-2.3ex]
\multirow{4}{*}{$\pt^{\PGg}$}
                               &20--80   & 1.10$^{+0.39}_{-0.38}$ & 0.0539 $\pm$ 0.0089   & 0.059 $\pm$ 0.021\\[2pt]
                               &80--120  & 1.96$^{+0.46}_{-0.43}$ & 0.0134 $\pm$ 0.0024   & 0.0264 $\pm$ 0.0060\\[2pt]
                               &120--200 & 1.51$^{+0.50}_{-0.46}$ & 0.0049 $\pm$ 0.0010   & 0.0074 $\pm$ 0.0024\\[2pt]
                               &200--400 & 1.29$^{+0.63}_{-0.57}$ & 0.00114 $\pm$ 0.00025 & 0.00147 $\pm$ 0.00068\\[\cmsTabSkip]
\multirow{4}{*}{$\pt^{\mathrm{j}_1}$}
                              &30--150  &  1.09$^{+0.59}_{-0.58}$ & 0.0176 $\pm$ 0.0028   & 0.019 $\pm$ 0.010\\[2pt]
                              &150--250 &  0.89$^{+0.34}_{-0.33}$ & 0.0149 $\pm$ 0.0026   & 0.0133 $\pm$ 0.0050\\[2pt]
                              &250--350 &  1.08$^{+0.50}_{-0.47}$ & 0.0052 $\pm$ 0.0010   & 0.0056 $\pm$ 0.0025\\[2pt]
                              &350--800 &  1.86$^{+0.63}_{-0.57}$ & 0.00059 $\pm$ 0.00014 & 0.00109 $\pm$ 0.0036\\[\cmsTabSkip] 
\multirow{4}{*}{$\pt^{\ell_1}$}
                            &20--80   & 1.60$^{+0.47}_{-0.47}$ & 0.0350 $\pm$ 0.0055   & 0.056 $\pm$ 0.016\\[2pt]
                            &80--120  & 1.01$^{+0.37}_{-0.35}$ & 0.0278 $\pm$ 0.0048   & 0.028 $\pm$ 0.010\\[2pt]
                            &120--200 & 0.30$^{+0.36}_{-0.34}$ & 0.0100 $\pm$ 0.0019   & 0.0030 $\pm$ 0.00035\\[2pt]
                            &200--400 & 0.97$^{+0.52}_{-0.47}$ & 0.00187 $\pm$ 0.00041 & 0.00188 $\pm$ 0.00092\\[2pt]          
\end{scotch}}
\label{tab:unfolding_results_1}
\vspace{0.5cm}
\end{table*}

\begin{table*}[htb]
\centering
\topcaption{The signal strengths and differential cross sections from SM expectation and fit calculated as part of the unfolding of 2D $m_{\mathrm{jj}}$-$\abs{\Delta\eta_{\mathrm{jj}}}$ observables for EW $\PZ\PGg$jj. The last bin includes overflow events.}
\cmsTable{
  \begin{scotch}{l l c c c}
  $\abs{\Delta\eta_{\mathrm{jj}}}$ bin & $m_{\mathrm{jj}}$ bin [\GeVns{}] & $\mu \pm \Delta\mu$ & Predicted $\rd^2\sigma/\rd m\,\rd\abs{\Delta\eta_{\mathrm{jj}}}$ [fb/\GeVns{}] & Observed $\rd^2\sigma/\rd m\,\rd\abs{\Delta\eta_{\mathrm{jj}}}$ [fb/\GeVns{}]\\ 
\hline \\  [-2.3ex]
[2.5, 4.5) & [500, 800)       &     1.25$^{+0.59}_{-0.58}$ & 0.00185 $\pm$ 0.00017   & 0.0023 $\pm$ 0.0011 \\[2pt]
[2.5, 4.5) & [800, 1200)      &     1.73$^{+0.43}_{-0.40}$ & 0.00096 $\pm$ 0.00014   & 0.00166 $\pm$ 0.00040 \\[2pt]
[2.5, 4.5) & [1200, 2000]     &     1.16$^{+0.34}_{-0.30}$ & 0.000322 $\pm$ 0.000065 & 0.00037 $\pm$ 0.00011\\[2pt]
[4.5, 6.0) & [500, 800)       &     4.3$^{+5.1}_{-4.8}$    & 0.000559 $\pm$ 0.000057 & 0.0024 $\pm$ 0.0028\\[2pt]
[4.5, 6.0) & [800, 1200)      &     1.3$^{+1.1}_{-1.1}$    & 0.00078 $\pm$ 0.00012   & 0.00104 $\pm$ 0.00086\\[2pt]
[4.5, 6.0) & [1200, 2000]     &     0.92$^{+0.28}_{-0.26}$ & 0.00069 $\pm$ 0.00016   & 0.00064 $\pm$ 0.00019\\[2pt]
[6.0, 6.5] & [500, 2000]      &     1.9$^{+1.5}_{-1.6}$    & 0.00060 $\pm$ 0.00016   & 0.00112 $\pm$ 0.00092\\[2pt]
  \end{scotch}}
  \label{tab:unfolding_results}
\end{table*}

\begin{figure*}[htbp!]
   \begin{center}
   \includegraphics[width=0.45\textwidth]{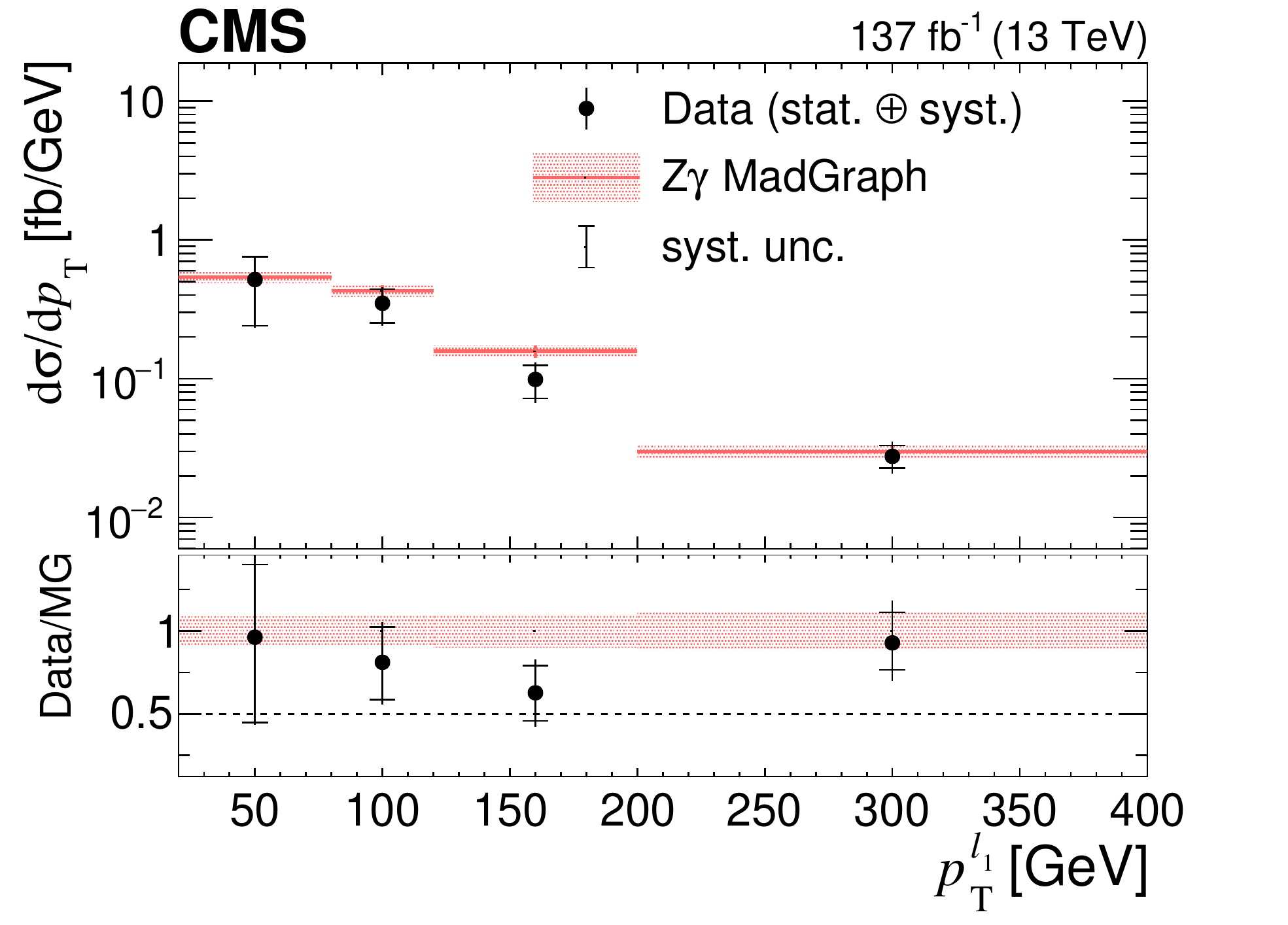}
   \includegraphics[width=0.45\textwidth]{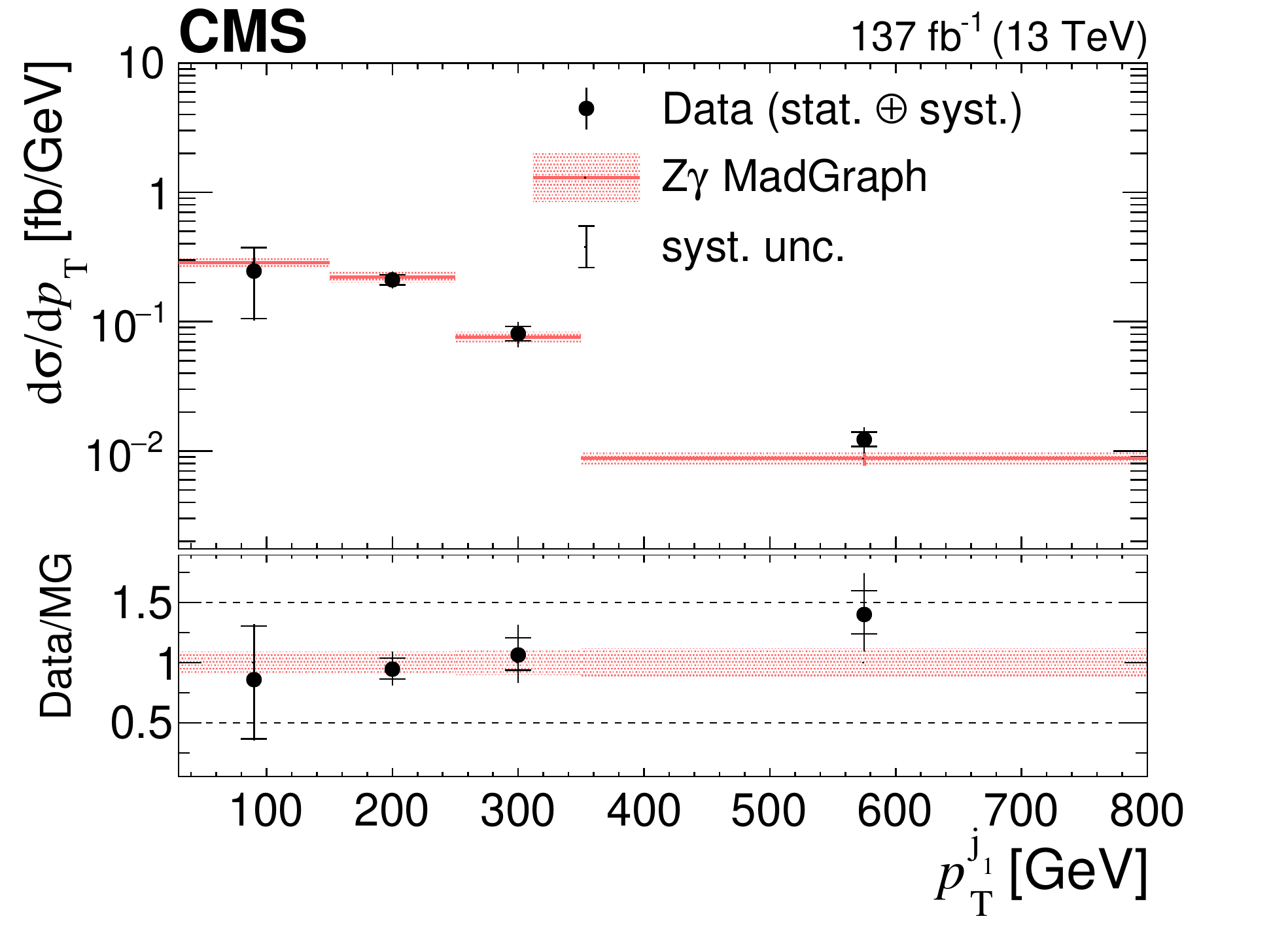}
   \\
   \includegraphics[width=0.45\textwidth]{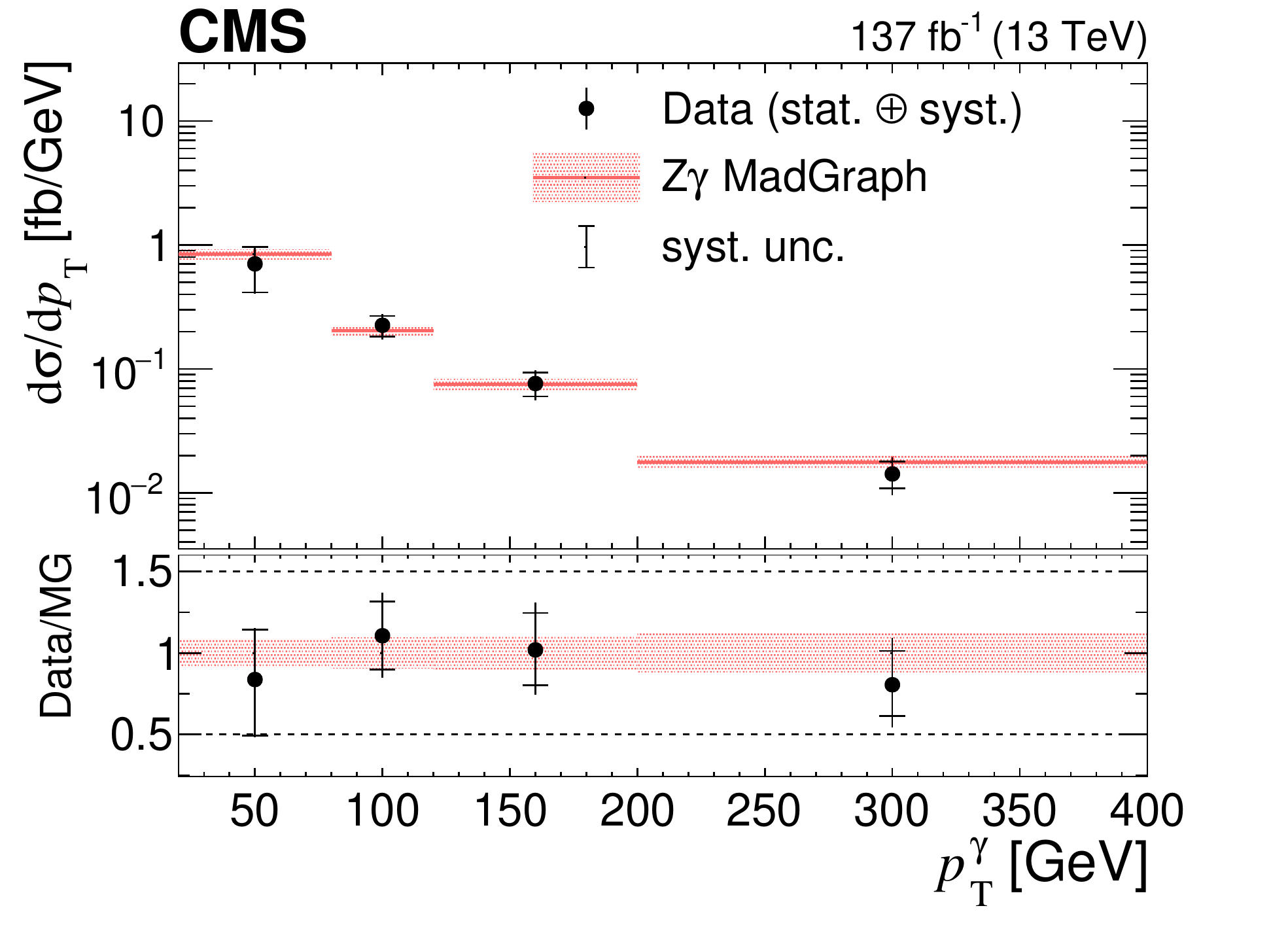}
    \includegraphics[width=0.45\textwidth]{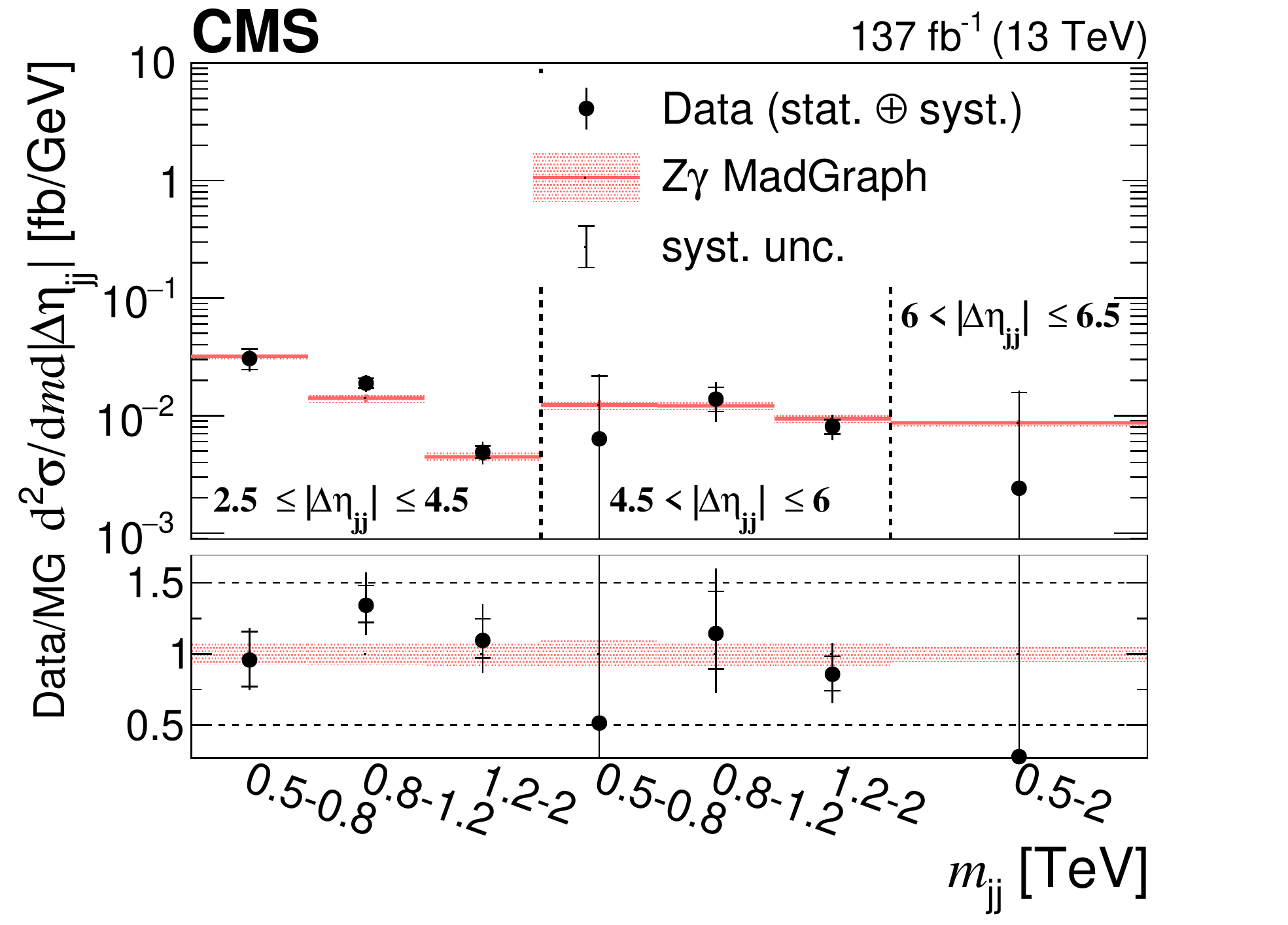}\\
   \caption{Unfolded differential cross section as a function of the leading lepton \pt, leading photon \pt, leading jet \pt, and $m_{\mathrm{jj}}$-$\abs{\Delta\eta_{\mathrm{jj}}}$ for EW+QCD $\PZ\PGg$jj. The black points with error bars represent the data and their statistical uncertainties, whereas the red bands represent the total theoretical uncertainties from the MG5 simulation. The last bin includes overflow events.}
   \label{fig:unfold_EW_QCD}
   \end{center}
\end{figure*}

A combined EW+QCD $\PZ\PGg$jj unfolded differential cross section is also measured in the same region using the same procedure, except that the control region is excluded. In this measurement, both the EW and QCD contributions are considered signal. The combined $\PZ\PGg$jj unfolded differential cross section is shown in Fig.~\ref{fig:unfold_EW_QCD}, and Tables~\ref{tab:unfolding_results_EW_QCD_1} and ~\ref{tab:unfolding_results_EW_QCD_2}. Within the uncertainties, the unfolded distributions agree well with the SM predictions.

\begin{table*}[hbtp!]
\centering
\topcaption{The signal strengths and differential cross sections from SM expectation and fit calculated as part of the unfolding of $\pt^{\PGg}$, $\pt^{\mathrm{j}_1}$, and $\pt^{\ell_1}$ observables for EW+QCD $\PZ\PGg$jj. The last bin includes overflow events.}
\cmsTable{
\begin{scotch}{ c l c c c  }
Variables & Bin [\GeVns{}] & $\mu \pm \Delta\mu$ & Predicted $\rd\sigma/\rd\pt$ [fb/\GeVns{}] & Observed $\rd\sigma/\rd\pt$ [fb/\GeVns{}]\\
\hline \\ [-2.3ex]
\multirow{4}{*}{$\pt^{\PGg}$}
                               &20--80   & 0.84$^{+0.32}_{-0.35}$ & 0.841 $\pm$ 0.075   & 0.70 $\pm$ 0.029\\[2pt]
                               &80--120  & 1.11$^{+0.27}_{-0.26}$ & 0.203 $\pm$ 0.019   & 0.225 $\pm$ 0.053\\[2pt]
                               &120--200 & 1.02$^{+0.29}_{-0.27}$ & 0.0747 $\pm$ 0.0076 & 0.076 $\pm$ 0.022\\[2pt]
                               &200--400 & 0.81$^{+0.29}_{-0.26}$ & 0.0176 $\pm$ 0.0021 & 0.0142 $\pm$ 0.0049\\[\cmsTabSkip] 
\multirow{4}{*}{$\pt^{\mathrm{j}_1}$}
                              &30--150  & 0.86$^{+0.46}_{-0.50}$ & 0.286 $\pm$ 0.025   & 0.25 $\pm$ 0.14\\[2pt]
                              &150--250 & 0.95$^{+0.15}_{-0.14}$ & 0.222 $\pm$ 0.020   & 0.210 $\pm$ 0.032\\[2pt]
                              &250--350 & 1.06$^{+0.25}_{-0.23}$ & 0.0759 $\pm$ 0.0077 & 0.081 $\pm$ 0.019\\[2pt]
                              &350--800 & 1.40$^{+0.35}_{-0.31}$ & 0.0087 $\pm$ 0.0010 & 0.0123 $\pm$ 0.0029\\[\cmsTabSkip]                                
\multirow{4}{*}{$\pt^{\ell_1}$}
                            &20--80   & 0.96$^{+0.45}_{-0.53}$ & 0.538 $\pm$ 0.046  & 0.52 $\pm$ 0.26\\[2pt]
                            &80--120  & 0.81$^{+0.24}_{-0.25}$ & 0.431 $\pm$ 0.040  & 0.35 $\pm$ 0.11\\[2pt]
                            &120--200 & 0.63$^{+0.20}_{-0.20}$ & 0.158 $\pm$ 0.016  & 0.099 $\pm$ 0.032\\[2pt]
                            &200--400 & 0.93$^{+0.26}_{-0.23}$ & 0.0297 $\pm$ 0.0034 & 0.0276 $\pm$ 0.0072\\[2pt]
\end{scotch}}
\label{tab:unfolding_results_EW_QCD_1}
\vspace{0.5cm}
\end{table*}

\begin{table*}[htb!]
\centering
\topcaption{The signal strengths and differential cross sections from SM expectation and fit calculated as part of the unfolding of 2D $m_{\mathrm{jj}}$-$\abs{\Delta\eta_{\mathrm{jj}}}$ observables for EW+QCD $\PZ\PGg$jj. The last bin includes overflow events.}
\cmsTable{
  \begin{scotch}{l l c c c}
  $\abs{\Delta\eta_{\mathrm{jj}}}$ bin & $m_{\mathrm{jj}}$ bin [\GeVns{}] & $\mu \pm \Delta\mu$ & Predicted $\rd^2\sigma/\rd m\,\rd\abs{\Delta\eta_{\mathrm{jj}}}$ [fb/\GeVns{}] & Observed $\rd^2\sigma/\rd m\,\rd\abs{\Delta\eta_{\mathrm{jj}}}$ [fb/\GeVns{}]\\
\hline\\ [-2.3ex]
[2.5, 4.5) & [500, 800)       & 0.96$^{+0.23}_{-0.21}$ & 0.0319 $\pm$ 0.0023   & 0.0306 $\pm$ 0.0070 \\[2pt]
[2.5, 4.5) & [800, 1200)      & 1.34$^{+0.23}_{-0.21}$ & 0.0140 $\pm$ 0.0011   & 0.0189 $\pm$ 0.0031\\[2pt]
[2.5, 4.5) & [1200, 2000]     & 1.09$^{+0.26}_{-0.23}$ & 0.00445 $\pm$ 0.00038 & 0.0049 $\pm$ 0.0010\\[2pt]
[4.5, 6.0) & [500, 800)       & 0.52$^{+1.3}_{-1.3}$   & 0.0123 $\pm$ 0.0012   & 0.006 $\pm$ 0.016\\[2pt]
[4.5, 6.0) & [800, 1200)      & 1.14$^{+0.46}_{-0.42}$ & 0.0121 $\pm$ 0.0010   & 0.0138 $\pm$ 0.0053\\[2pt]
[4.5, 6.0) & [1200, 2000]     & 0.86$^{+0.22}_{-0.20}$ & 0.00942 $\pm$ 0.00076 & 0.0081 $\pm$ 0.0020\\[2pt]
[6.0, 6.5] & [500, 2000]      & 0.3$^{+1.6}_{-1.6}$    & 0.00864 $\pm$ 0.00049 & 0.0024 $\pm$ 0.0014\\[2pt]
  \end{scotch}}
  \label{tab:unfolding_results_EW_QCD_2}
\end{table*}

\subsection{Limits on anomalous quartic gauge couplings}
In an EFT approach to BSM physics, dimension-8 operators are constructed from contractions of the covariant derivative of the Higgs doublet and the charged and neutral field strength tensors associated with gauge bosons. Assuming the $SU(2)\times U(1)$ symmetry of the EW gauge field, nine independent charge-conjugate and parity-conserving dimension-8 operators are constructed~\cite{Eboli:2006wa}. The operators affecting the $\PZ\PGg$jj channel can be divided into those containing an $SU(2)$ field strength, the $U(1)$ field strength, and the covariant derivative of the Higgs doublet, $\mathcal{L}_{\mathrm{M_{0-5,7}}}$, and those containing only the two field strengths, $\mathcal{L}_{\mathrm{T_{0-2,5-9}}}$. The coefficient of the operator $\mathcal{L}_{\mathrm{X_{Y}}}$ is denoted by $F_{\mathrm{XY}}/\Lambda^4$, where $\Lambda$ is the unknown scale of BSM physics.

\begin{figure}[htb!]
   \centering
      \includegraphics[width=\cmsFigWidth]{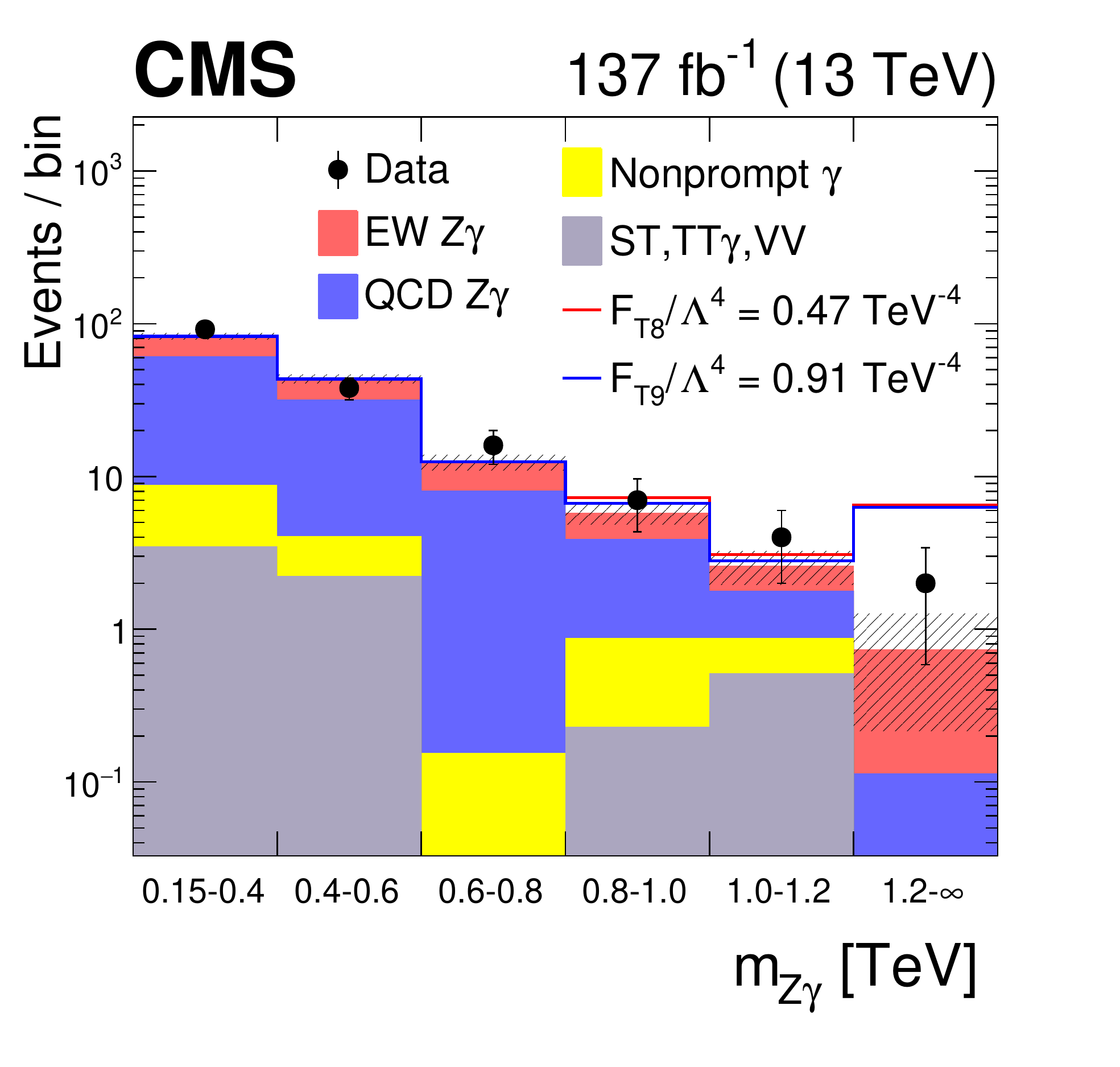}
      \caption{The $\mzg$ distribution for events satisfying the aQGC region selection, which is used to set constraints on the anomalous coupling parameters. The bins of $\mzg$ are [150, 400, 600, 800, 1000, 1200, 2000]\GeV, where the last bin includes overflow events. The red line represents a nonzero $F_{\mathrm{T8}}$ value and the blue line represents a nonzero $F_{\mathrm{T9}}$ value, which would significantly enhance the yields at high $\mzg$. The black points with error bars represent the data and their statistical uncertainties, whereas the hatched bands represent the statistical uncertainties in the SM predictions.}
      \label{fig:aqgc_yields}
\end{figure}

The effects of the aQGCs in addition to the SM EW $\PZ\PGg$ process, as well as interference between the EW and QCD-induced processes, are simulated as described in Sec~\ref{sec:samples}. The contribution from aQGCs enhances the production of events at large $\PZ\PGg$ mass, so the $\mzg$ distribution is used to extract limits on the aQGC parameters. To obtain the prediction for the signal as a function of the value of the aQGC parameter, a quadratic function is used to fit the ratio of aQGC and SM yields in each bin of $\mzg$ in the aQGC region defined in Sec~\ref{sec:event_selection}. From Fig.~\ref{fig:aqgc_yields}, no statistically significant excess of events relative to the SM prediction is observed. 

The likelihood function is the product of Poisson distributions and a normal constraining term with nuisance parameters representing the sources of systematic uncertainties in any given bin. The final likelihood function is the product of the likelihood functions of the electron and muon channels. This test statistic,
\begin{linenomath}
\begin{equation}\label{likelihoodformula}
t_{\alpha_\mathrm{test}} = -2 \ln \frac{{\cal L}(\alpha_\text{test},{\hat{\hat{\vec{\theta}}}})}{\mathcal{L}(\hat{\alpha},\hat{\vec{\theta}})},
\end{equation}
\end{linenomath}
is essentially the same test statistic as in Sec~\ref{sec7} except that the $\alpha_{\mathrm{test}}$ represents the aQGC point being tested. The symbol $\vec{\theta}$ represents a vector of nuisance parameters assumed to follow log-normal distributions. The parameter $\hat{\hat{\vec{\theta}}}$ corresponds to the maximum of the likelihood function at the point $\alpha_{\mathrm{test}}$. The $\hat{\alpha}$ and $\hat{\vec{\theta}}$ parameters correspond to the global maximum of the likelihood function.

This test statistic is assumed to follow a $\chi^2$ distribution~\cite{1943Wald:wilks1938}. It is therefore possible to extract the limits immediately from twice the difference in the log-likelihood function $2\Delta\mathrm{NLL} = t_{\alpha_{\mathrm{test}}}$~\cite{Khachatryan:2014jba}. The observed and expected 95\% confidence level limits for the coefficients, shown in Table~\ref{tab:VBS_aQGC}, are obtained by varying the coefficients of one operator at a time and setting all other anomalous couplings to zero. The statistical uncertainty is dominant in the limits setting. The unitarity bound is defined as the scattering energy at which the aQGC coupling strength is set equal to the observed limit that would result in a scattering amplitude that violates unitarity~\cite{Almeida_2020}. These results provide the most stringent limit to date on the aQGC parameter $F_{\mathrm{T9}}/\Lambda^4$.

\begin{table*}[htb!]
\centering
\topcaption{The expected and observed limits on the aQGC parameters at 95\% confidence level. The last column presents the scattering energy values for which the amplitude would violate unitarity for the observed value of the aQGC parameter. All coupling parameter limits are set in $\TeV^{-4}$, whereas the unitarity bounds are in TeV.}
  \begin{scotch}{cccccc}
 Coupling & Exp. lower & Exp. upper & Obs. lower & Obs. upper & Unitarity bound\\
  \hline
      $F_\mathrm{M0}/\Lambda^{4}$ &  $-$12.5 & 12.8 & $-$15.8 & 16.0  & 1.3   \\
      $F_\mathrm{M1}/\Lambda^{4}$ &  $-$28.1 & 27.0 & $-$35.0 & 34.7  & 1.5  \\
      $F_\mathrm{M2}/\Lambda^{4}$ &  $-$5.21 & 5.12 & $-$6.55 & 6.49  & 1.5  \\
      $F_\mathrm{M3}/\Lambda^{4}$ &  $-$10.2 & 10.3 & $-$13.0 & 13.0  & 1.8  \\
      $F_\mathrm{M4}/\Lambda^{4}$ &  $-$10.2 & 10.2 & $-$13.0 & 12.7  & 1.7  \\
      $F_\mathrm{M5}/\Lambda^{4}$ &  $-$17.6 & 16.8 & $-$22.2 & 21.3  & 1.7  \\
      $F_\mathrm{M7}/\Lambda^{4}$ &  $-$44.7 & 45.0 & $-$56.6 & 55.9  & 1.6  \\
      $F_\mathrm{T0}/\Lambda^{4}$ &  $-$0.52 & 0.44 & $-$0.64 & 0.57  & 1.9  \\
      $F_\mathrm{T1}/\Lambda^{4}$ &  $-$0.65 & 0.63 & $-$0.81 & 0.90  & 2.0  \\
      $F_\mathrm{T2}/\Lambda^{4}$ &  $-$1.36 & 1.21 & $-$1.68 & 1.54  & 1.9  \\
      $F_\mathrm{T5}/\Lambda^{4}$ &  $-$0.45 & 0.52 & $-$0.58 & 0.64  & 2.2  \\
      $F_\mathrm{T6}/\Lambda^{4}$ &  $-$1.02 & 1.07 & $-$1.30 & 1.33  & 2.0  \\
      $F_\mathrm{T7}/\Lambda^{4}$ &  $-$1.67 & 1.97 & $-$2.15 & 2.43  & 2.2  \\
      $F_\mathrm{T8}/\Lambda^{4}$ &  $-$0.36 & 0.36 & $-$0.47 & 0.47  & 1.8  \\
      $F_\mathrm{T9}/\Lambda^{4}$ &  $-$0.72 & 0.72 & $-$0.91 & 0.91  & 1.9  \\
  \end{scotch}
  \label{tab:VBS_aQGC}
\end{table*}

\section{Summary}
\label{sec:summary}
This paper presents the first observation of the electroweak (EW) production of a {\PZ} boson, a photon, and two jets ($\PZ\PGg$jj) in $\sqrt{s}=13\TeV$ proton-proton collisions recorded with the CMS detector in 2016--2018 corresponding to an integrated luminosity of 137\fbinv. Events were selected by requiring two opposite-sign leptons with the same flavor from the decay of a {\PZ} boson, one identified photon, and two jets that have a large separation in pseudorapidity and a large dijet mass. The measured cross section in the fiducial volume defined in Table~\ref{tab:selections} for EW $\PZ\PGg$jj production is $5.21\pm0.52\stat\pm0.56\syst\unit{fb}=5.21\pm0.76\unit{fb}$, and the fiducial cross section of EW and QCD-induced production is $ 14.7\pm0.80\stat\pm1.26\syst\unit{fb}=14.7\pm1.53\unit{fb}$. Both the observed and expected signal significances are well in excess of 5 standard deviations. Differential cross sections for EW and EW+QCD are measured for several observables and compared to standard model predictions computed at leading order. Within the uncertainties, the measurements agree with the predictions. Constraints are set on the effective field theory dimension-8 operators $\mathrm{M}_{0}$ to $\mathrm{M}_{5}$, $\mathrm{M}_{7}$, $\mathrm{T}_{0}$ to $\mathrm{T}_{2}$, and $\mathrm{T}_{5}$ to $\mathrm{T}_{9}$, giving rise to anomalous quartic gauge couplings. These constraints are either competitive with or more stringent than those previously obtained.

\begin{acknowledgments}
   We congratulate our colleagues in the CERN accelerator departments for the excellent performance of the LHC and thank the technical and administrative staffs at CERN and at other CMS institutes for their contributions to the success of the CMS effort. In addition, we gratefully acknowledge the computing centers and personnel of the Worldwide LHC Computing Grid and other centers for delivering so effectively the computing infrastructure essential to our analyses. Finally, we acknowledge the enduring support for the construction and operation of the LHC, the CMS detector, and the supporting computing infrastructure provided by the following funding agencies: BMBWF and FWF (Austria); FNRS and FWO (Belgium); CNPq, CAPES, FAPERJ, FAPERGS, and FAPESP (Brazil); MES (Bulgaria); CERN; CAS, MoST, and NSFC (China); MINCIENCIAS (Colombia); MSES and CSF (Croatia); RIF (Cyprus); SENESCYT (Ecuador); MoER, ERC PUT and ERDF (Estonia); Academy of Finland, MEC, and HIP (Finland); CEA and CNRS/IN2P3 (France); BMBF, DFG, and HGF (Germany); GSRT (Greece); NKFIA (Hungary); DAE and DST (India); IPM (Iran); SFI (Ireland); INFN (Italy); MSIP and NRF (Republic of Korea); MES (Latvia); LAS (Lithuania); MOE and UM (Malaysia); BUAP, CINVESTAV, CONACYT, LNS, SEP, and UASLP-FAI (Mexico); MOS (Montenegro); MBIE (New Zealand); PAEC (Pakistan); MSHE and NSC (Poland); FCT (Portugal); JINR (Dubna); MON, RosAtom, RAS, RFBR, and NRC KI (Russia); MESTD (Serbia); SEIDI, CPAN, PCTI, and FEDER (Spain); MOSTR (Sri Lanka); Swiss Funding Agencies (Switzerland); MST (Taipei); ThEPCenter, IPST, STAR, and NSTDA (Thailand); TUBITAK and TAEK (Turkey); NASU (Ukraine); STFC (United Kingdom); DOE and NSF (USA).
   
   \hyphenation{Rachada-pisek} Individuals have received support from the Marie-Curie program and the European Research Council and Horizon 2020 Grant, contracts No.\ 675440, 724704, 752730, 765710 and 824093 (European Union); the Leventis Foundation; the Alfred P.\ Sloan Foundation; the Alexander von Humboldt Foundation; the Belgian Federal Science Policy Office; the Fonds pour la Formation \`a la Recherche dans l'Industrie et dans l'Agriculture (FRIA-Belgium); the Agentschap voor Innovatie door Wetenschap en Technologie (IWT-Belgium); the F.R.S.-FNRS and FWO (Belgium) under the ``Excellence of Science -- EOS" -- be.h project n.\ 30820817; the Beijing Municipal Science \& Technology Commission, No. Z191100007219010; the Ministry of Education, Youth and Sports (MEYS) of the Czech Republic; the Deutsche Forschungsgemeinschaft (DFG), under Germany's Excellence Strategy -- EXC 2121 ``Quantum Universe" -- 390833306, and under project No. 400140256 - GRK2497; the Lend\"ulet (``Momentum") Program and the J\'anos Bolyai Research Scholarship of the Hungarian Academy of Sciences, the New National Excellence Program \'UNKP, the NKFIA research grants 123842, 123959, 124845, 124850, 125105, 128713, 128786, and 129058 (Hungary); the Council of Science and Industrial Research, India; the Latvian Council of Science; the Ministry of Science and Higher Education and the National Science Center, contracts Opus 2014/15/B/ST2/03998 and 2015/19/B/ST2/02861 (Poland); the National Priorities Research Program by Qatar National Research Fund; the Ministry of Science and Higher Education, project no. 0723-2020-0041 (Russia); the Programa Estatal de Fomento de la Investigaci{\'o}n Cient{\'i}fica y T{\'e}cnica de Excelencia Mar\'{\i}a de Maeztu, grant MDM-2015-0509 and the Programa Severo Ochoa del Principado de Asturias; the Thalis and Aristeia programs cofinanced by EU-ESF and the Greek NSRF; the Rachadapisek Sompot Fund for Postdoctoral Fellowship, Chulalongkorn University and the Chulalongkorn Academic into Its 2nd Century Project Advancement Project (Thailand); the Kavli Foundation; the Nvidia Corporation; the SuperMicro Corporation; the Welch Foundation, contract C-1845; and the Weston Havens Foundation (USA).\end{acknowledgments}

\bibliography{auto_generated}

\providecommand{\href}[2]{#2}\begingroup\raggedright\begin{thebibliography}{10}%
\makeatletter
\providecommand{\hrefCMSnoop }[0]{\@secondoftwo}%
\makeatother
\providecommand{\doi}{\texttt{doi:}\begingroup \urlstyle{tt}\Url}

\bibitem{Aad:2012tfa}
\hrefCMSnoop {}{{ATLAS Collaboration}, ``{Observation of a new particle in the
  search for the standard model Higgs boson with the ATLAS detector at the
  LHC}'',} \textit{ Phys. Lett. B} \textbf{ 716} (2012) 1,
  \href{http://dx.doi.org/10.1016/j.physletb.2012.08.020}{\doi{10.1016/j.physletb.2012.08.020}},
  \href{http://www.arXiv.org/abs/1207.7214}{\texttt{arXiv:1207.7214}}.

\bibitem{Chatrchyan:2012ufa}
\hrefCMSnoop {}{{CMS Collaboration}, ``{Observation of a new boson at a mass of
  125 GeV with the CMS experiment at the LHC}'',} \textit{ Phys. Lett. B}
  \textbf{ 716} (2012) 30,
  \href{http://dx.doi.org/10.1016/j.physletb.2012.08.021}{\doi{10.1016/j.physletb.2012.08.021}},
\href{http://www.arXiv.org/abs/1207.7235}{\texttt{arXiv:1207.7235}}.
%%CITATION = ARXIV:1207.7235;%%.

\bibitem{Chatrchyan:2012ufa_long}
\hrefCMSnoop {}{{CMS Collaboration}, ``{Observation of a new boson with mass
  near 125 GeV in $\Pp\Pp$ collisions at $\sqrt{s} =$ 7 and 8 TeV}'',} \textit{
  JHEP} \textbf{ 06} (2013) 081,
  \href{http://dx.doi.org/10.1007/JHEP06(2013)081}{\doi{10.1007/JHEP06(2013)081}},
\href{http://www.arXiv.org/abs/1303.4571}{\texttt{arXiv:1303.4571}}.
%%CITATION = ARXIV:1303.4571;%%.

\bibitem{higgs_measure_2016}
\hrefCMSnoop {}{{ATLAS, CMS} Collaboration, ``{Measurements of the Higgs boson
  production and decay rates and constraints on its couplings from a combined
  ATLAS and CMS analysis of the LHC pp collision data at $ \sqrt{s}=7 $ and 8
  TeV}'',} \textit{ JHEP} \textbf{ 08} (2016) 045,
  \href{http://dx.doi.org/10.1007/JHEP08(2016)045}{\doi{10.1007/JHEP08(2016)045}},
  \href{http://www.arXiv.org/abs/1606.02266}{\texttt{arXiv:1606.02266}}.

\bibitem{higgs_measure_2019}
\hrefCMSnoop {}{{CMS Collaboration}, ``{Combined measurements of Higgs boson
  couplings in proton-proton collisions at $\sqrt{s} =$ 13 TeV}'',} \textit{
  Eur. Phys. J. C} \textbf{ 79} (2019) 421,
  \href{http://dx.doi.org/10.1140/epjc/s10052-019-6909-y}{\doi{10.1140/epjc/s10052-019-6909-y}},
\href{http://www.arXiv.org/abs/1809.10733}{\texttt{arXiv:1809.10733}}.
%%CITATION = ARXIV:1809.10733;%%.

\bibitem{Eboli:2006wa}
\hrefCMSnoop {}{O.~J.~P. {\'{E}}boli, M.~C. Gonzalez-Garcia, and J.~K.
  Mizukoshi, ``{$\Pp\Pp$$\rightarrow$ jje$^\pm$$\mu^\pm$$\nu\nu$ and
  jje$^\pm\mu^\mp\nu\nu$ at $\mathcal{O}$($\alpha^6_{\rm em}$) and
  $\mathcal{O}$($\alpha_{\rm em}^4\, \alpha_{\rm s}^2$) for the study of the
  quartic electroweak gauge boson vertex at CERN LHC}'',} \textit{ Phys. Rev.
  D} \textbf{ 74} (2006) 073005,
  \href{http://dx.doi.org/10.1103/PhysRevD.74.073005}{\doi{10.1103/PhysRevD.74.073005}},
\href{http://www.arXiv.org/abs/hep-ph/0606118}{\texttt{arXiv:hep-ph/0606118}}.
%%CITATION = HEP-PH/0606118;%%.

\bibitem{aTGC_WV}
\hrefCMSnoop {}{{CMS Collaboration}, ``{Search for anomalous triple gauge
  couplings in WW and WZ production in lepton + jet events in proton-proton
  collisions at $\sqrt{s} =$ 13 TeV}'',} \textit{ JHEP} \textbf{ 12} (2019)
  062,
  \href{http://dx.doi.org/10.1007/JHEP12(2019)062}{\doi{10.1007/JHEP12(2019)062}},
  \href{http://www.arXiv.org/abs/1907.08354}{\texttt{arXiv:1907.08354}}.

\bibitem{Aad:2019wpb}
\hrefCMSnoop {}{{ATLAS Collaboration}, ``{Evidence for electroweak production
  of two jets in association with a {\PZ}$\gamma$ pair in pp collisions at
  $\sqrt{s} = 13$ TeV with the ATLAS detector}'',} \textit{ Phys. Lett. B}
  \textbf{ 803} (2020) 135341,
  \href{http://dx.doi.org/10.1016/j.physletb.2020.135341}{\doi{10.1016/j.physletb.2020.135341}},
  \href{http://www.arXiv.org/abs/1910.09503}{\texttt{arXiv:1910.09503}}.

\bibitem{Sirunyan:2020tlu}
\hrefCMSnoop {}{{CMS Collaboration}, ``{Measurement of the cross section for
  electroweak production of a Z boson, a photon and two jets in proton-proton
  collisions at $\sqrt{s} =$ 13 TeV and constraints on anomalous quartic
  couplings}'',} \textit{ JHEP} \textbf{ 06} (2020) 076,
  \href{http://dx.doi.org/10.1007/JHEP06(2020)076}{\doi{10.1007/JHEP06(2020)076}},
  \href{http://www.arXiv.org/abs/2002.09902}{\texttt{arXiv:2002.09902}}.

\bibitem{Chatrchyan:2008zzk}
\hrefCMSnoop {}{{CMS Collaboration}, ``The {CMS} experiment at the {CERN}
  {LHC}'',} \textit{ JINST} \textbf{ 3} (2008) S08004,
  \href{http://dx.doi.org/10.1088/1748-0221/3/08/S08004}{\doi{10.1088/1748-0221/3/08/S08004}}.

\bibitem{trigger_system}
\hrefCMSnoop {}{{CMS Collaboration}, ``The {CMS} trigger system'',} \textit{
  JINST} \textbf{ 12} (2017) P01020,
  \href{http://dx.doi.org/10.1088/1748-0221/12/01/P01020}{\doi{10.1088/1748-0221/12/01/P01020}},
  \href{http://www.arXiv.org/abs/1609.02366}{\texttt{arXiv:1609.02366}}.

\bibitem{MGatNLO}
J.~Alwall\hrefCMSnoop {}{ {et~al.}, ``The automated computation of tree-level
  and next-to-leading order differential cross sections, and their matching to
  parton shower simulations'',} \textit{ JHEP} \textbf{ 07} (2014) 079,
  \href{http://dx.doi.org/10.1007/JHEP07(2014)079}{\doi{10.1007/JHEP07(2014)079}},
\href{http://www.arXiv.org/abs/1405.0301}{\texttt{arXiv:1405.0301}}.
%%CITATION = ARXIV:1405.0301;%%.

\bibitem{POWHEG}
\hrefCMSnoop {}{T.~Melia, P.~Nason, R.~Rontsch, and G.~Zanderighi,
  ``{$\text{W}^{+}\text{W}^{-}$, {\PW}{\PZ} and {\PZ}{\PZ} production in the
  POWHEG BOX}'',} \textit{ JHEP} \textbf{ 11} (2011) 078,
  \href{http://dx.doi.org/10.1007/JHEP11(2011)078}{\doi{10.1007/JHEP11(2011)078}},
\href{http://www.arXiv.org/abs/1107.5051}{\texttt{arXiv:1107.5051}}.
%%CITATION = ARXIV:1107.5051;%%.

\bibitem{POWHEG1}
\hrefCMSnoop {}{P.~Nason, ``{A new method for combining NLO QCD with shower
  Monte Carlo algorithms}'',} \textit{ JHEP} \textbf{ 11} (2004) 040,
  \href{http://dx.doi.org/10.1088/1126-6708/2004/11/040}{\doi{10.1088/1126-6708/2004/11/040}},
\href{http://www.arXiv.org/abs/hep-ph/0409146}{\texttt{arXiv:hep-ph/0409146}}.
%%CITATION = HEP-PH/0409146;%%.

\bibitem{POWHEG2}
\hrefCMSnoop {}{S.~Frixione, P.~Nason, and C.~Oleari, ``{Matching NLO QCD
  computations with parton shower simulations: the POWHEG method}'',} \textit{
  JHEP} \textbf{ 11} (2007) 070,
  \href{http://dx.doi.org/10.1088/1126-6708/2007/11/070}{\doi{10.1088/1126-6708/2007/11/070}},
\href{http://www.arXiv.org/abs/0709.2092}{\texttt{arXiv:0709.2092}}.
%%CITATION = 0709.2092;%%.

\bibitem{POWHEG3}
\hrefCMSnoop {}{S.~Alioli, P.~Nason, C.~Oleari, and E.~Re, ``{A general
  framework for implementing NLO calculations in shower Monte Carlo programs:
  the POWHEG BOX}'',} \textit{ JHEP} \textbf{ 06} (2010) 043,
  \href{http://dx.doi.org/10.1007/JHEP06(2010)043}{\doi{10.1007/JHEP06(2010)043}},
\href{http://www.arXiv.org/abs/1002.2581}{\texttt{arXiv:1002.2581}}.
%%CITATION = ARXIV:1002.2581;%%.

\bibitem{Sjostrand:2014zea}
T.~Sj{\"o}strand\hrefCMSnoop {}{ {et~al.}, ``An introduction to {PYTHIA}
  8.2'',} \textit{ Comput. Phys. Commun.} \textbf{ 191} (2015) 159,
  \href{http://dx.doi.org/10.1016/j.cpc.2015.01.024}{\doi{10.1016/j.cpc.2015.01.024}},
\href{http://www.arXiv.org/abs/1410.3012}{\texttt{arXiv:1410.3012}}.
%%CITATION = ARXIV:1410.3012;%%.

\bibitem{Frederix:2012ps}
\hrefCMSnoop {}{R.~Frederix and S.~Frixione, ``{Merging meets matching in
  MC@NLO}'',} \textit{ JHEP} \textbf{ 12} (2012) 061,
  \href{http://dx.doi.org/10.1007/JHEP12(2012)061}{\doi{10.1007/JHEP12(2012)061}},
\href{http://www.arXiv.org/abs/1209.6215}{\texttt{arXiv:1209.6215}}.
%%CITATION = ARXIV:1209.6215;%%.

\bibitem{mg5:reweight}
\hrefCMSnoop {}{O.~Mattelaer, ``{On the maximal use of Monte Carlo samples:
  re-weighting events at NLO accuracy}'',} \textit{ Eur. Phys. J. C} \textbf{
  76} (2016) 674,
  \href{http://dx.doi.org/10.1140/epjc/s10052-016-4533-7}{\doi{10.1140/epjc/s10052-016-4533-7}},
\href{http://www.arXiv.org/abs/1607.00763}{\texttt{arXiv:1607.00763}}.
%%CITATION = ARXIV:1607.00763;%%.

\bibitem{nnpdf}
\hrefCMSnoop {}{{NNPDF} Collaboration, ``{Parton distributions for the LHC run
  II}'',} \textit{ JHEP} \textbf{ 04} (2015) 040,
  \href{http://dx.doi.org/10.1007/JHEP04(2015)040}{\doi{10.1007/JHEP04(2015)040}},
  \href{http://www.arXiv.org/abs/1410.8849}{\texttt{arXiv:1410.8849}}.

\bibitem{nnpdf_new}
\hrefCMSnoop {}{{NNPDF} Collaboration, ``Parton distributions from
  high-precision collider data'',} \textit{ Eur. Phys. J. C} \textbf{ 77}
  (2017) 663,
  \href{http://dx.doi.org/10.1140/epjc/s10052-017-5199-5}{\doi{10.1140/epjc/s10052-017-5199-5}},
  \href{http://www.arXiv.org/abs/1706.00428}{\texttt{arXiv:1706.00428}}.

\bibitem{Skands:2014pea}
\hrefCMSnoop {}{P.~Skands, S.~Carrazza, and J.~Rojo, ``Tuning {PYTHIA} 8.1: the
  {Monash} 2013 tune'',} \textit{ Eur. Phys. J. C} \textbf{ 74} (2014) 3024,
  \href{http://dx.doi.org/10.1140/epjc/s10052-014-3024-y}{\doi{10.1140/epjc/s10052-014-3024-y}},
  \href{http://www.arXiv.org/abs/1404.5630}{\texttt{arXiv:1404.5630}}.

\bibitem{Khachatryan:2015pea}
\hrefCMSnoop {}{{CMS Collaboration}, ``Event generator tunes obtained from
  underlying event and multiparton scattering measurements'',} \textit{ Eur.
  Phys. J. C} \textbf{ 76} (2016) 155,
  \href{http://dx.doi.org/10.1140/epjc/s10052-016-3988-x}{\doi{10.1140/epjc/s10052-016-3988-x}},
\href{http://www.arXiv.org/abs/1512.00815}{\texttt{arXiv:1512.00815}}.
%%CITATION = ARXIV:1512.00815;%%.

\bibitem{geant4}
\hrefCMSnoop {}{{GEANT4} Collaboration, ``{\GEANTfour}--a simulation
  toolkit'',} \textit{ Nucl. Instrum. Meth. A} \textbf{ 506} (2003) 250,
\href{http://dx.doi.org/10.1016/S0168-9002(03)01368-8}{\doi{10.1016/S0168-9002(03)01368-8}}.
%%CITATION = NUIMA,A506,250;%%.

\bibitem{geant4_2}
\hrefCMSnoop {}{{GEANT4} Collaboration, ``{GEANT4} developments and
  applications'',} \textit{ IEEE Trans. Nucl. Sci.} \textbf{ 53} (2006) 270,
  \href{http://dx.doi.org/10.1109/TNS.2006.869826}{\doi{10.1109/TNS.2006.869826}}.

\bibitem{Sirunyan:2017ulk}
\hrefCMSnoop {}{{CMS Collaboration}, ``{Particle-flow reconstruction and global
  event description with the CMS detector}'',} \textit{ JINST} \textbf{ 12}
  (2017) P10003,
  \href{http://dx.doi.org/10.1088/1748-0221/12/10/P10003}{\doi{10.1088/1748-0221/12/10/P10003}},
\href{http://www.arXiv.org/abs/1706.04965}{\texttt{arXiv:1706.04965}}.
%%CITATION = ARXIV:1706.04965;%%.

\bibitem{antikt}
\hrefCMSnoop {}{M.~Cacciari, G.~P. Salam, and G.~Soyez, ``{The anti-\kt jet
  clustering algorithm}'',} \textit{ JHEP} \textbf{ 04} (2008) 063,
  \href{http://dx.doi.org/10.1088/1126-6708/2008/04/063}{\doi{10.1088/1126-6708/2008/04/063}},
\href{http://www.arXiv.org/abs/0802.1189}{\texttt{arXiv:0802.1189}}.
%%CITATION = 0802.1189;%%.

\bibitem{Cacciari:fastjet1}
\hrefCMSnoop {}{M.~Cacciari, G.~P. Salam, and G.~Soyez, ``{FastJet} user
  manual'',} \textit{ Eur. Phys. J. C} \textbf{ 72} (2012) 1896,
  \href{http://dx.doi.org/10.1140/epjc/s10052-012-1896-2}{\doi{10.1140/epjc/s10052-012-1896-2}},
  \href{http://www.arXiv.org/abs/1111.6097}{\texttt{arXiv:1111.6097}}.

\bibitem{cmscollaboration2020electron}
\hrefCMSnoop {}{{CMS Collaboration}, ``{Electron and photon reconstruction and
  identification with the CMS experiment at the CERN LHC}'',} \textit{ JINST}
  \textbf{ 16} (2021) P05014,
  \href{http://dx.doi.org/10.1088/1748-0221/16/05/P05014}{\doi{10.1088/1748-0221/16/05/P05014}},
  \href{http://www.arXiv.org/abs/2012.06888}{\texttt{arXiv:2012.06888}}.

\bibitem{Khachatryan:2015hwa}
\hrefCMSnoop {}{{CMS Collaboration}, ``Performance of electron reconstruction
  and selection with the {CMS} detector in proton-proton collisions at
  {$\sqrt{s} = 8\TeV$}'',} \textit{ JINST} \textbf{ 10} (2015) P06005,
  \href{http://dx.doi.org/10.1088/1748-0221/10/06/P06005}{\doi{10.1088/1748-0221/10/06/P06005}},
\href{http://www.arXiv.org/abs/1502.02701}{\texttt{arXiv:1502.02701}}.
%%CITATION = ARXIV:1502.02701;%%.

\bibitem{muon_13tev}
\hrefCMSnoop {}{{CMS Collaboration}, ``{Performance of the CMS muon detector
  and muon reconstruction with proton-proton collisions at $\sqrt{s}=$ 13
  TeV}'',} \textit{ JINST} \textbf{ 13} (2018) P06015,
  \href{http://dx.doi.org/10.1088/1748-0221/13/06/P06015}{\doi{10.1088/1748-0221/13/06/P06015}},
  \href{http://www.arXiv.org/abs/1804.04528}{\texttt{arXiv:1804.04528}}.

\bibitem{jetarea_method}
\hrefCMSnoop {}{M.~Cacciari and G.~P. Salam, ``{Pileup subtraction using jet
  areas}'',} \textit{ Phys. Lett. B} \textbf{ 659} (2008) 119,
  \href{http://dx.doi.org/10.1016/j.physletb.2007.09.077}{\doi{10.1016/j.physletb.2007.09.077}},
  \href{http://www.arXiv.org/abs/0707.1378}{\texttt{arXiv:0707.1378}}.

\bibitem{tagandprobe}
\hrefCMSnoop {}{{CMS Collaboration}, ``{Measurement of the inclusive {\PW} and
  {\PZ} production cross sections in $\Pp\Pp$ collisions at $\sqrt{s} =$ 7 TeV
  with the CMS experiment}'',} \textit{ JHEP} \textbf{ 10} (2011) 132,
  \href{http://dx.doi.org/10.1007/JHEP10(2011)132}{\doi{10.1007/JHEP10(2011)132}},
  \href{http://www.arXiv.org/abs/1107.4789}{\texttt{arXiv:1107.4789}}.

\bibitem{electron_7tev}
\hrefCMSnoop {}{{CMS Collaboration}, ``{Energy calibration and resolution of
  the CMS electromagnetic calorimeter in $\Pp\Pp$ collision at $\sqrt{s}=$ 7
  TeV}'',} \textit{ JINST} \textbf{ 8} (2013) P09009,
  \href{http://dx.doi.org/10.1088/1748-0221/8/09/P09009}{\doi{10.1088/1748-0221/8/09/P09009}},
  \href{http://www.arXiv.org/abs/1306.2016}{\texttt{arXiv:1306.2016}}.

\bibitem{photon_8tev}
\hrefCMSnoop {}{{CMS Collaboration}, ``Performance of photon reconstruction and
  identification with the {CMS} detector in proton-proton collisions at
  {$\sqrt{s} = 8\TeV$}'',} \textit{ JINST} \textbf{ 10} (2015) P08010,
  \href{http://dx.doi.org/10.1088/1748-0221/10/08/P08010}{\doi{10.1088/1748-0221/10/08/P08010}},
\href{http://www.arXiv.org/abs/1502.02702}{\texttt{arXiv:1502.02702}}.
%%CITATION = ARXIV:1502.02702;%%.

\bibitem{jer}
\hrefCMSnoop {}{{CMS Collaboration}, ``{Jet energy scale and resolution in the
  CMS experiment in $\Pp\Pp$ collisions at 8 TeV}'',} \textit{ JINST} \textbf{
  12} (2017) P02014,
  \href{http://dx.doi.org/10.1088/1748-0221/12/02/P02014}{\doi{10.1088/1748-0221/12/02/P02014}},
  \href{http://www.arXiv.org/abs/1607.03663}{\texttt{arXiv:1607.03663}}.

\bibitem{puId}
\hrefCMSnoop {}{{CMS Collaboration}, ``{Pileup mitigation at CMS in 13 TeV
  data}'',} \textit{ JINST} \textbf{ 15} (2020) P09018,
  \href{http://dx.doi.org/10.1088/1748-0221/15/09/P09018}{\doi{10.1088/1748-0221/15/09/P09018}},
  \href{http://www.arXiv.org/abs/2003.00503}{\texttt{arXiv:2003.00503}}.

\bibitem{zeppenfeld}
\hrefCMSnoop {}{D.~Rainwater, R.~Szalapski, and D.~Zeppenfeld, ``{Probing color
  singlet exchange in {\PZ}+2-jet events at the CERN LHC}'',} \textit{ Phys.
  Rev. D} \textbf{ 54} (1996) 6680,
  \href{http://dx.doi.org/10.1103/PhysRevD.54.6680}{\doi{10.1103/PhysRevD.54.6680}},
  \href{http://www.arXiv.org/abs/hep-ph/9605444}{\texttt{arXiv:hep-ph/9605444}}.

\bibitem{STW_1}
\hrefCMSnoop {}{N.~Kidonakis, ``{Two-loop soft anomalous dimensions for single
  top quark associated production with a $W^{-}$ or $H^{-}$}'',} \textit{ Phys.
  Rev. D} \textbf{ 82} (2010) 054018,
  \href{http://dx.doi.org/10.1103/PhysRevD.82.054018}{\doi{10.1103/PhysRevD.82.054018}},
\href{http://www.arXiv.org/abs/1005.4451}{\texttt{arXiv:1005.4451}}.
%%CITATION = ARXIV:1005.4451;%%.

\bibitem{Butterworth:2015oua}
\hrefCMSnoop {}{J.~Butterworth {et~al.}, ``{PDF4LHC} recommendations for {LHC
  Run II}'',} \textit{ J. Phys. G} \textbf{ 43} (2016) 023001,
  \href{http://dx.doi.org/10.1088/0954-3899/43/2/023001}{\doi{10.1088/0954-3899/43/2/023001}},
\href{http://www.arXiv.org/abs/1510.03865}{\texttt{arXiv:1510.03865}}.
%%CITATION = ARXIV:1510.03865;%%.

\bibitem{minibais}
\hrefCMSnoop {}{{CMS Collaboration}, ``{Measurement of the inelastic
  proton-proton cross section at $ \sqrt{s}=13 $ TeV}'',} \textit{ JHEP}
  \textbf{ 07} (2018) 161,
  \href{http://dx.doi.org/10.1007/JHEP07(2018)161}{\doi{10.1007/JHEP07(2018)161}},
  \href{http://www.arXiv.org/abs/1802.02613}{\texttt{arXiv:1802.02613}}.

\bibitem{CMS-LUM-17-003}
\hrefCMSnoop {}{{CMS Collaboration}, ``{Precision luminosity measurement in
  proton-proton collisions at $\sqrt{s} =$ 13 TeV in 2015 and 2016 at CMS}'',}
  \textit{ Eur. Phys. J. C} \textbf{ 81} (2021) 800,
  \href{http://dx.doi.org/10.1140/epjc/s10052-021-09538-2}{\doi{10.1140/epjc/s10052-021-09538-2}},
  \href{http://www.arXiv.org/abs/2104.01927}{\texttt{arXiv:2104.01927}}.

\bibitem{CMS:2018elu}
\href {https://cds.cern.ch/record/2621960}{{CMS Collaboration}, ``{CMS
  luminosity measurement for the 2017 data-taking period at $\sqrt{s} =
  13~\mathrm{TeV}$}'',} {Physics Analysis Summary} CMS-PAS-LUM-17-004, CERN,
  Geneva, 2018.

\bibitem{CMS:2019jhq}
\href {https://cds.cern.ch/record/2676164}{{CMS Collaboration}, ``{CMS
  luminosity measurement for the 2018 data-taking period at $\sqrt{s} =
  13~\mathrm{TeV}$}'',} {Physics Analysis Summary} CMS-PAS-LUM-18-002, CERN,
  Geneva, 2019.

\bibitem{pvalue}
\hrefCMSnoop {}{L.~Demortier, ``{P} values and nuisance parameters'',} in
  \textit{ Statistical issues for {LHC} physics. {Proceedings, Workshop,
  PHYSTAT-LHC, Geneva, Switzerland, June} 27-29, 2007}.
\newblock 2008.
\newblock
\href{http://dx.doi.org/10.5170/CERN-2008-001}{\doi{10.5170/CERN-2008-001}}.
%%CITATION = CERN-2008-001;%%.

\bibitem{Cowan:2010js}
\hrefCMSnoop {}{G.~Cowan, K.~Cranmer, E.~Gross, and O.~Vitells, ``Asymptotic
  formulae for likelihood-based tests of new physics'',} \textit{ Eur. Phys. J.
  C} \textbf{ 71} (2011) 1554,
  \href{http://dx.doi.org/10.1140/epjc/s10052-011-1554-0}{\doi{10.1140/epjc/s10052-011-1554-0}},
  \href{http://www.arXiv.org/abs/1007.1727}{\texttt{arXiv:1007.1727}}.
[Erratum: \DOI{10.1140/epjc/s10052-013-2501-z}].
%%CITATION = ARXIV:1007.1727;%%.

\bibitem{doi:10.1137/1.9780898718836.ch4}
\hrefCMSnoop {}{P.~C. Hansen, ``Computational aspects: Regularization
  methods'',} in \textit{ Discrete inverse problems: insight and algorithms}.
\newblock SIAM, 2010.
\newblock
  \href{http://dx.doi.org/10.1137/1.9780898718836.ch4}{\doi{10.1137/1.9780898718836.ch4}}.

\bibitem{1943Wald:wilks1938}
\hrefCMSnoop {}{S.~S. Wilks, ``{The large-sample distribution of the likelihood
  ratio for testing composite hypotheses}'',} \textit{ Ann. Math. Statist}
  \textbf{ 9} (1938) 60,
  \href{http://dx.doi.org/10.1214/aoms/1177732360}{\doi{10.1214/aoms/1177732360}}.

\bibitem{Khachatryan:2014jba}
\hrefCMSnoop {}{{CMS Collaboration}, ``{Precise determination of the mass of
  the Higgs boson and tests of compatibility of its couplings with the standard
  model predictions using proton collisions at 7 and 8 TeV}'',} \textit{ Eur.
  Phys. J. C} \textbf{ 75} (2015) 212,
  \href{http://dx.doi.org/10.1140/epjc/s10052-015-3351-7}{\doi{10.1140/epjc/s10052-015-3351-7}},
\href{http://www.arXiv.org/abs/1412.8662}{\texttt{arXiv:1412.8662}}.
%%CITATION = ARXIV:1412.8662;%%.

\bibitem{Almeida_2020}
\hrefCMSnoop {}{E.~d.~S. Almeida, O.~J.~P. \'Eboli, and M.~C.
  Gonzalez\textendash{}Garcia, ``{Unitarity constraints on anomalous quartic
  couplings}'',} \textit{ Phys. Rev. D} \textbf{ 101} (2020) 113003,
  \href{http://dx.doi.org/10.1103/PhysRevD.101.113003}{\doi{10.1103/PhysRevD.101.113003}},
  \href{http://www.arXiv.org/abs/2004.05174}{\texttt{arXiv:2004.05174}}.

\end{thebibliography}\endgroup

\cleardoublepage \appendix\section{The CMS Collaboration \label{app:collab}}\begin{sloppypar}\hyphenpenalty=5000\widowpenalty=500\clubpenalty=5000\vskip\cmsinstskip
\textbf{Yerevan Physics Institute, Yerevan, Armenia}\\*[0pt]
A.~Tumasyan
\vskip\cmsinstskip
\textbf{Institut f\"{u}r Hochenergiephysik, Wien, Austria}\\*[0pt]
W.~Adam, J.W.~Andrejkovic, T.~Bergauer, S.~Chatterjee, M.~Dragicevic, A.~Escalante~Del~Valle, R.~Fr\"{u}hwirth\cmsAuthorMark{1}, M.~Jeitler\cmsAuthorMark{1}, N.~Krammer, L.~Lechner, D.~Liko, I.~Mikulec, P.~Paulitsch, F.M.~Pitters, J.~Schieck\cmsAuthorMark{1}, R.~Sch\"{o}fbeck, M.~Spanring, S.~Templ, W.~Waltenberger, C.-E.~Wulz\cmsAuthorMark{1}
% ...
\vskip\cmsinstskip
% ...
1:  Also at TU Wien, Wien, Austria\\
% ...
\end{sloppypar}
%%% END EDITABLE REGION %%%
% skeleton_end
\end{document}